\documentclass[apj]{emulateapj}
\usepackage{graphics}
\usepackage{natbib}
\newcommand{\as}{$^{\prime\prime}$}
\newcommand{\am}{$^{\prime}$}

\begin{document}

\title{Broadband X-ray Imaging and Spectroscopy of the Crab Nebula and Pulsar with \textit{NuSTAR} }
\author{Kristin K. Madsen$^1$, Stephen Reynolds$^2$, Fiona Harrison$^1$, Hongjun An$^3$, Steven Boggs$^4$, Finn E. Christensen$^5$, William W. Craig$^4$, Chris L. Fryer$^6$, Brian W. Grefenstette$^1$, Charles J. Hailey$^7$, Craig Markwardt$^8$, Melania Nynka$^7$, Daniel Stern$^9$, Andreas Zoglauer$^4$ and William Zhang$^8$}
\affiliation{$^1$ Cahill Center for Astronomy and Astrophysics, California Institute of Technology, Pasadena, CA 91125, USA\\
$^2$ Physics Department, NC State University, Raleigh, NC 27695, USA\\
$^3$ Department of Physics, McGill University, Montreal, Quebec, H3A 2T8, Canada\\
$^4$ Space Sciences Laboratory, University of California, Berkeley, CA 94720, USA\\
$^5$ DTU Space, National Space Institute, Technical University of Denmark, Elektronvej 327, DK-2800 Lyngby, Denmark\\
$^6$ CCS-2, Los Alamos National Laboratory, Livermore, CA 94550, USA\\
$^7$ Columbia Astrophysics Laboratory, Columbia University, New York 10027, USA\\
$^8$ Goddard Space Flight Center, Greenbelt, MD 20771, USA\\
$^9$ Jet Propulsion Laboratory, California Institute of Technology, Pasadena, CA 91109, USA\\
}

\begin{abstract}
We present broadband (3 -- 78 keV) \textit{NuSTAR}  X-ray imaging and spectroscopy of the Crab nebula and pulsar. We show that while the phase-averaged and spatially integrated nebula + pulsar spectrum is a power-law in this energy band, spatially resolved spectroscopy of the nebula finds a break at $\sim$9 keV in the spectral photon index of the torus structure with a steepening characterized by $\Delta\Gamma\sim0.25$. We also confirm a previously reported steepening in the pulsed spectrum, and quantify it with a broken power-law with break energy at $\sim$12 keV and $\Delta\Gamma\sim0.27$. We present spectral maps of the inner 100\as\ of the remnant and measure the size of the nebula as a function of energy in seven bands. These results find that the rate of shrinkage with energy of the torus size can be fitted by a power-law with an index of $\gamma = 0.094\pm 0.018$,  consistent with the predictions of Kennel and Coroniti (1984). The change in size is more rapid in the NW direction, coinciding with the counter-jet where we find the index to be a factor of two larger. \textit{NuSTAR} observed the Crab during the latter part of a $\gamma$-ray flare, but found no increase in flux in the 3 - 78 keV energy band.
\end{abstract}


\section{Introduction}
The Crab is the prototypical Pulsar Wind Nebula (PWN), characterized by a center-filled synchrotron nebula that is powered by a magnetized wind of charged particles emanating from a centrally located pulsar formed during the supernova explosion \citep{Weiler1978}. Due to its brightness, proximity of $\sim$2 kpc, and well-known explosion date of 1054 AD, the Crab is the best studied object of its kind, and the literature is rich with hundreds of publications and reviews describing its properties across all energy bands (see e.g. \citet{Hester2008} and \citet{Buhler2014} for a recent review). 

Detailed images from the \textit{Hubble Space Telescope} \citep{Hester1995} and \textit{Chandra} \citep{Weisskopf2000} have revealed the nebula's morphological complexities. In the optical band the remnant measures $\sim$3\am\ across its longest axis with thermal filaments composed of ejecta from the explosion confining the synchrotron nebula. In X-rays the remnant is considerably smaller and shows both torus and jet structures. The symmetry axis is tilted at about $27^\circ$ to the plane of the sky with the NW edge closer to the observer \citep{Ng2004}. The jet emerges toward the observer to the SE, and a less collimated structure (the counter-jet) extends away to the NW. \citet{Ng2004} fit simple models to the morphology to obtain a radial flow velocity through the torus of $0.550 \pm 0.001 c$, and at these velocities relativistic beaming brightens the nearside (NW) of the torus. 

The wind likely terminates at a shock zone about 10\as\ from the pulsar \citep{Weisskopf2000}, and the post shock particles and magnetic fields propagate non-relativistically either by diffusion \citep{Gratton1972,Wilson19722} or advection \citep{Rees1974,KC19841,KC19842} to the edge of the remnant, emitting their energy as synchrotron radiation. Both the process of particle acceleration in relativistic shocks and the transport of particles and fields downstream, occur in other astrophysical settings such as gamma-ray bursts and jets in active galaxies, and studying them in a relatively nearby spatially resolved source, may have application beyond the understanding of the Crab and PWNe in general.

From radio to TeV, the emission from the Crab nebula is non-thermal and peaks in the optical through X-ray. At radio wavelengths the nebula's integrated emission is a power-law spectrum with index ($S_\nu \propto \nu^\alpha$) $\alpha = -0.299 \pm 0.009$ \citep{Baars1977}. In the optical, the synchotron spectrum is steeper with a gradual turnover occurring somewhere between 10 and 1000 $\mu$m \citep{Marsden1984}. In the X-ray band the Crab nebula+pulsar photon index is $\Gamma \sim 2.1$ and further softens above 100\,keV to $\sim$2.23 \citep{Jourdain2009}. The Crab nebula component alone likewise softens above 100\,keV to $\sim$2.14 up to 300\,keV \citep{Pravdo1997} and softens further to $2.227 \pm 0.013$ between 0.75 and 30\,MeV \citep{Kuiper2001} where it continues to soften until 700\,MeV, beyond which the inverse Compton component sets in \citep{Meyer2010}. The pulsed spectrum follows a different spectral energy distribution (SED) with a more complex spectral evolution \citep{Kuiper2001,Weisskopf2011,Kirsch2006}.

In GeV $\gamma$-rays the Crab has been observed to flare rapidly \citep{Tavani2011,Abdo2011} roughly once a year for a duration of $\sim$10 days, with a flux increase of a factor of $\sim$5. The origins of these flaring episodes are currently not understood, but due to the rapidity of the flares and the fast cooling time of synchrotron radiation, the flare emission is presumed to be of synchrotron origin rather than inverse Compton or bremsstrahlung \citep{Abdo2011}.

No single model has yet successfully been devised to explain the properties of the nebula across all energy bands. The ratio of wind flow times to the particle radiative lifetimes is a strong model dependent parameter tied to the mechanism of energy transport and is observable as an energy-dependent nebular size. The prediction of both diffusion and advection models is a decrease in size of the nebula with increasing photon energy due to higher-energy electrons dying out sooner than lower-energy ones. In X-rays the expectation is for a diffusion-driven nebula to be smaller than one that is advection dominated. \citet{Ku1976} first collected broadband evidence confirming shrinkage of the Crab nebula, and this limited data set pointed towards a combination of diffusive and advective electron transport.

The spatial dependence of the broad-band spectrum provides another observational constraint on models. Detailed \textit{Chandra} observations \citep{Mori2004} show that the torus spectrum is roughly uniform with a steepening at the edge, in line with predictions made by \citet{KC19841,KC19842}, and indicates that for X-ray-emitting particles, the transport in the Crab nebula appears to be dominated by advection rather than diffusion.

\begin{table}
\caption{Observations Log}
\centering
\begin{tabular}{lcccl}
\hline
Obs ID & Date & \footnote{Effective exposure time corrected for dead-time.}Exposure & off-axis angle & Comment \\
& & seconds & arcminutes &  \\
\hline 
\hline
10013022002\footnote{Subset used for deconvolution analysis.} & 2012-09-20 & 2592 & 1.5/2.0 & Crab22 \\ 
10013022004$^b$ & 2012-09-21 & 2347 &  1.5/2.0 & Crab22 \\ 
10013022006$^b$ & 2012-09-21 & 2587 &  1.5/2.0 & Crab22 \\
10013031002$^b$ & 2012-10-25 & 2507 &  2.0/2.6 & Crab31 \\  
10013032002$^b$ & 2012-11-04 & 2595 &  1.2/1.3 & Crab32 \\
10013034002 & 2013-02-14 & 988  & 0.8/1.5 & Crab34 \\  
10013034004 & 2013-02-14 & 5720 & 0.7/0.9 & Crab34 \\ 
10013034005 & 2013-02-15 & 5968 & 0.9/0.6 & Crab34 \\
10013037002 & 2013-04-03 & 2679 & 2.0/2.4 & Crab37 \\
10013037004 & 2013-04-04 & 2796 & 2.9/3.5 & Crab37 \\ 
10013037006 & 2013-04-05 & 2944 & 2.9/3.5 & Crab37 \\
10013037008 & 2013-04-18 & 2814 & 3.0/3.7 & Crab37 \\
80001022002$^b$ & 2013-03-09 & 3917 & 1.5/1.8 & CrabToo \\
10002001002$^b$ & 2013-09-02 & 2608 & 1.72/2.09 & CrabSci \\
10002001004$^b$ & 2013-09-03 & 2386 & 1.73/2.26 & CrabSci \\
10002001006$^b$ & 2013-11-11 & 14260 & 1.08/1.28 & CrabSci \\
\hline
\end{tabular}
\label{obsid}
\end{table} 

Above 10 keV, observations with the ability to spatially resolve the Crab nebula have been limited to one-dimensional scanning techniques and lunar occultation, which have limited signal to noise and lack true two-dimensional imaging \citep{Pelling1987}. The \textit{NuSTAR} high-energy X-ray focusing mission has the ability to make sensitive imaging observations of diffuse sources above 10~keV for the first time. \textit{NuSTAR's} two co-aligned telescopes operate from 3 -- 78 keV, and have an imaging resolution of 18\as\ FWHM and $\sim1$\am\ Half Power Diameter \citep{Harrison2013}. \textit{NuSTAR} can image the Crab above 10 keV with sufficient resolution to investigate the spectral and spatial properties of the nebula.

In Section \ref{phaseaveraged} we present the global properties of the pulsar and nebula. In Section \ref{flaring} we discuss the \textit{NuSTAR} observations during the $\gamma$-ray flaring on the 9th of March 2013. In \ref{phaseresolved} we present phase resolved spectroscopy of the Crab in the energy range 3 -- 78 keV, tracing the spectrum of the pulsar as a function of its phase. In Section \ref{spatialspectral} we present spectral maps of the inner 100\as\ of the nebula, using pulse-off phase intervals, and in Section \ref{size} we present deconvolved \textit{NuSTAR} images to investigate the physical size of the Crab as a function of energy. In the discussion section we review and summarize the new findings.

\section{Observations and Data Reduction}

\textit{NuSTAR} has two co-aligned telescopes with corresponding focal planes FPMA and FPMB.  Each focal plane consists of four hybrid hard X-ray pixel detectors with a total field of view of 12.5\am$\times 12.5$\am. The detectors are denoted by numbers 0 through 3, with the optical axis placed in the inner corner of detector 0. \textit{NuSTAR} observed the Crab multiple times throughout 2012 and 2013 as part of the instrument calibration campaign. The observations span a wide range of off-axis angles, some of which were deemed too far off-axis for the analysis presented here, and Table \ref{obsid} lists the chosen subset. All of these observations have the pulsar within 4\am\ of the optical axis and located on detector 0, with a total elapsed live-time corrected exposure of 59.7 ks. The accuracy of the absolute source location can vary due to the thermal environment with $3\sigma$ offsets typically on the order of 8\as\ \citep{Harrison2013}, but because of the very peaked pulsar emission it is straightforward to register all observations to the pulsar location.

Despite being a bright target ($\sim$250 cts s$^{-1}$ per FPM with a dead time of $\sim50\%$), no special processing or pile-up corrections are needed since the focal plane detectors have a triggered readout with a short (1~$\mu$s) preamplifier shaping time \citet{Harrison2013}. The data were reduced using the NuSTARDAS pipeline version v1.3.0 and CALDB version 20131223 with all standard settings. We extracted spectra using standard scripts provided by the NuSTARDAS pipeline.

The background for the Crab only becomes important at energies above 60\,keV where the internal detector background dominates over sky background. Since the internal background varies from detector to detector, ideally one would extract background from the same detector as the source. Unfortunately the brightness and extent of the Crab precludes extracting a local background from the same detector. We therefore simulated the backgrounds using the \textit{nuskybgd} tool \citep{Wik2014} and created a master background for each observation for a 200\as\ extraction radius. To test for fluctuations in the background, we ran numerous realizations fitting a typical representative spectrum. We found that the fits were not sensitive to the background fluctuations.

In the text and tables all errors are reported at 90\% confidence, and all fits are performed with XSPEC\footnote{http://heasarc.gsfc.nasa.gov/xanadu/xspec/} using Cash statistics \citep{Cash1979} unless otherwise stated. Spectra shown are rebinned for display purposes only.

\begin{figure}
\begin{center}
\includegraphics[width=0.47\textwidth]{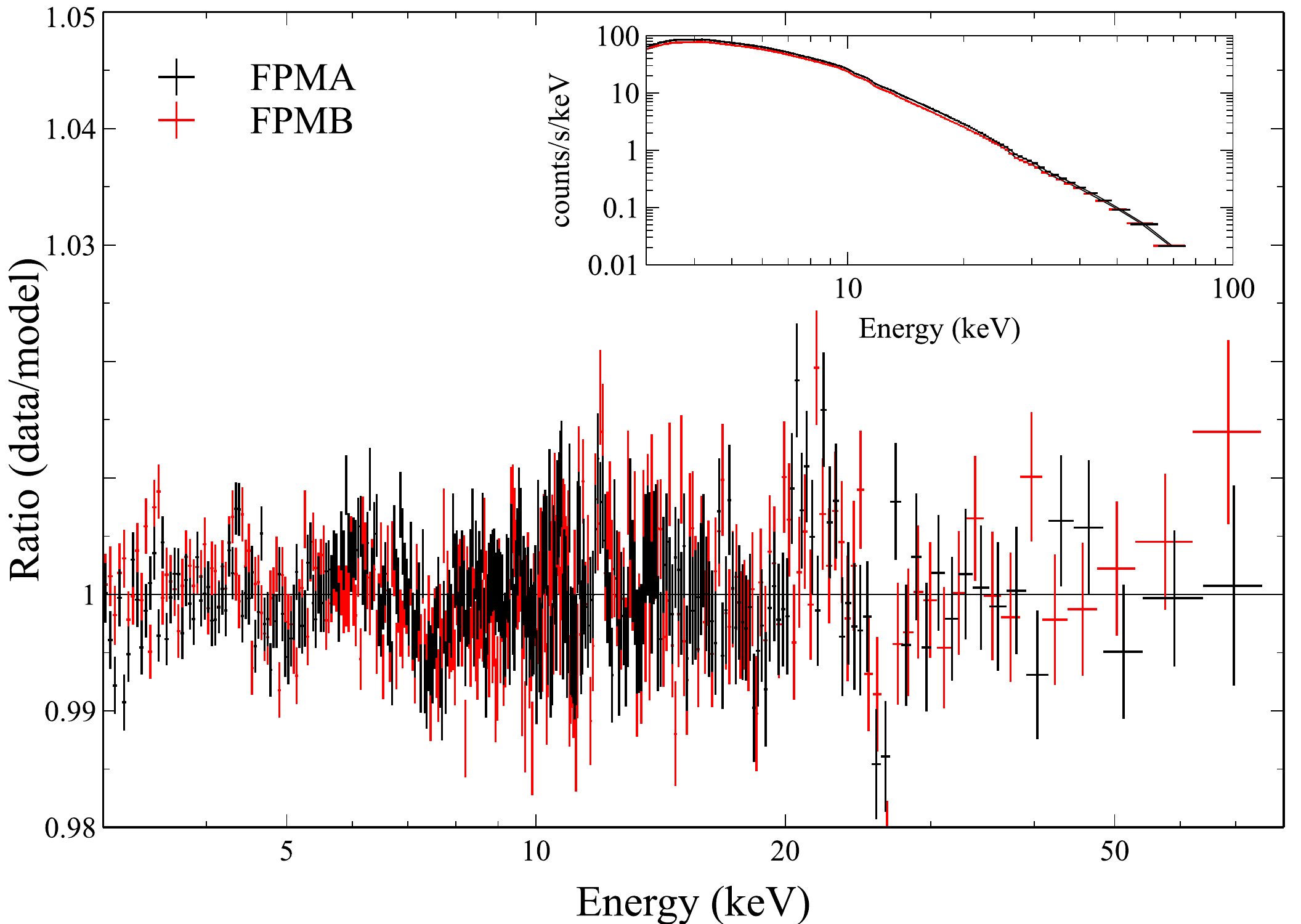}
\end{center}
\caption{Ratio of the data to the best-fit model for all the Crab observations listed in Table \ref{obsid}. The structured residuals, such as those around 20 -- 25\,keV are related to the calibration process (piece-wise linear spline interpolation), but are small ($\sim$2\%), typically broad and at energies not to be mistaken for emission lines.}
\label{canonical}
\end{figure}

\begin{figure}
\begin{center}
\includegraphics[width=0.47\textwidth]{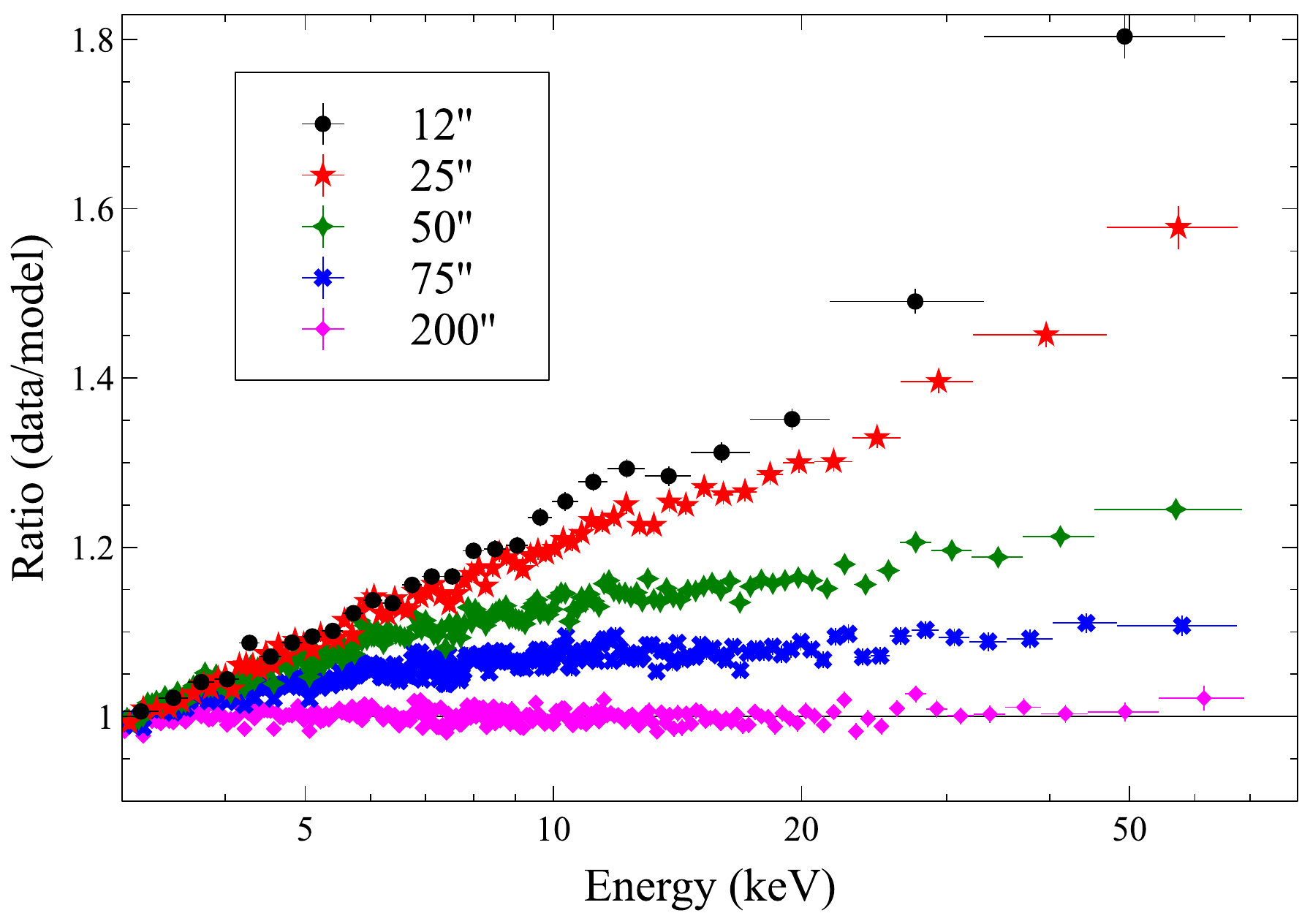}
\end{center}
\caption{Spectra and fit of FPMA assuming a power-law index of $\Gamma=2.1$. The normalizations of the curves have been scaled to illustrates how the index of the Crab softens with increasing extraction radius until at an extraction radius of 200\as\ the power-law index of 2.1 is recovered.}
\label{cumulative}
\end{figure}

\section{Data analysis}
\subsection{Phase Averaged Spectroscopy}\label{phaseaveraged}
The spatially integrated spectrum of the Crab nebula + pulsar in the 1 -- 100\,keV X-ray band has been well-described by a power-law with photon index $\Gamma\sim 2.1$ (\textit{RXTE, BeppoSAX, EXOSAT, INTEGRAL}/JEM\_X) \citep{Kirsch2005}. Above 100\,keV the hard X-ray instruments (\textit{INTEGRAL}/SPI/ISGRI, \textit{CGRO}) measure a softer index of $\Gamma \sim$2.20 -- 2.25, and below 10\,keV instruments with CCD detectors a harder spectrum. This hardening in CCD X-ray instruments comes about from photon pile-up, and although models exist for these instruments to deal with the pile-up, the Crab usually still challenges the models and requires special non-standard reductions. In addition, it is common practice for these instruments to excise the piled-up regions, which removes part of the integrated spectrum and thus can technically no longer be directly compared to other instruments where this excision has not occurred. The non piled-up instruments covering the 1 -- 100\,keV band agree on a photon index $\Gamma = 2.1 \pm 0.02$ and have not measured any curvature in the Crab spectrum across this band.

Over the 16 years {\em RXTE } was operational and regularly monitoring the Crab, the spectral index was seen to vary by a peak-to-peak variation of $\Delta \Gamma \sim 0.025$ \citep{Shaposhnikov2012}, perhaps due to magnetosonic waves in the nebula (e.g., \citet{Spitkovsky2004}). This variation is consistent with the observed spread between instruments, but is slow and on average over the 16 years the Crab remained at $\Gamma = 2.1$. Because the average index covers several instruments and the deviations from it are small, \textit{NuSTAR} has calibrated the effective area against a Crab index of $\Gamma = 2.1$.

A total of 39 Crab observations, spanning off-axis angles from 0 -- 7\am, went into adjusting the effective area ancillary response files (ARF) of \textit{NuSTAR}. This was done using a piece-wise linear spline interpolation as a function of energy and off-axis angle. Cross-calibration campaigns on 3C\,273 and PKS2155-304 have been used to confirm that the ARF adjustments have not introduced a systematic offset, and well-known power-law sources, such as CenA, to confirm they still appear as power-law sources. We used the quasar 3C\,273 to calibrate the N$_H$ column and derived a value of N$_H= (2 \pm 2) \times 10^{21}$ cm$^{-2}$ for the Crab. At 3\,keV \textit{NuSTAR} is only very marginally sensitive to N$_H$ columns of $10^{21}$ cm$^{-2}$, which is the reason behind the large error on the N$_H$ column. We have frozen the column to the above for all fits, employing Wilms abundances \citep{Wilms2000} and Verner cross-sections \citep{Verner1996}. An extensive discussion on the choice of N$_H$ and effective area calibration can be found in \citet{Madsen2014}. 

The Crab flux has been registered in multiple instruments to decline over a 2 year period of $\sim 7\%$ \citep{Wilson2011} across the 15 -- 50\,keV band. This corresponds to a decay in the flux of 3.5\% per year over the period it has been observed. We set the Crab normalization to 8.5 to optimize (and minimize) the cross-calibration constants between concurrent X-ray observatories (\textit{Chandra, Swift, Suzaku, XMM-Newton}). In the followed, however, the absolute flux or a variation of it, has no influence on the results.

We combine all the observations from Table \ref{obsid}, excluding those denoted by ``Crab34" that have the central region of the source falling on the gap between detectors in a way that complicates flux correction. Figure \ref{canonical} shows the ratio of data to the best-fit power-law model $\Gamma = 2.0963 \pm 0.0004$ and cross normalization between FPMA and FPMB of 1.002. The structured residuals, such as those around 20 -- 25\,keV are related to the calibration process and are small ($\sim$2\%), typically broad and at energies not to be mistaken for emission lines. 

It is important to note that in the subsequent analysis, we are focusing on changes relative to the above measured spectrum. The data quality is high, and we can measure slight changes within the spectrum with great accuracy. This is not to be mistaken for our knowledge of the \textit{absolute} value of the spectral index. Individual fits to the 39 Crab observations yields an average measured photon-index of $\Gamma = 2.1$ with a $1\sigma$ spread of $\pm 0.01$. However, this error on our absolute measurement is not relevant to our analysis since we are only concerned with relative changes.

To illustrate the range of spectral change in the Crab's integrated spectrum as a function of radius, we extract all events within circles of radius: 12\as, 25\as, 50\as, 75\as, 150\as\ and 200\as\ centered on the Crab pulsar and subtract the simulated master background.  Figure \ref{cumulative} shows the ratio of these spectra to the canonical power-law model $\Gamma=2.1$, where the relative normalizations have been set equal at 3\,keV. As already discussed, this progressive softening of the index is predicted by theory and has been measured at all energies (up to 10\,keV) where it is possible to spatially resolve the Crab.

\begin{figure}
\begin{center}
\includegraphics[width=0.55\textwidth, viewport=100 50 600 700]{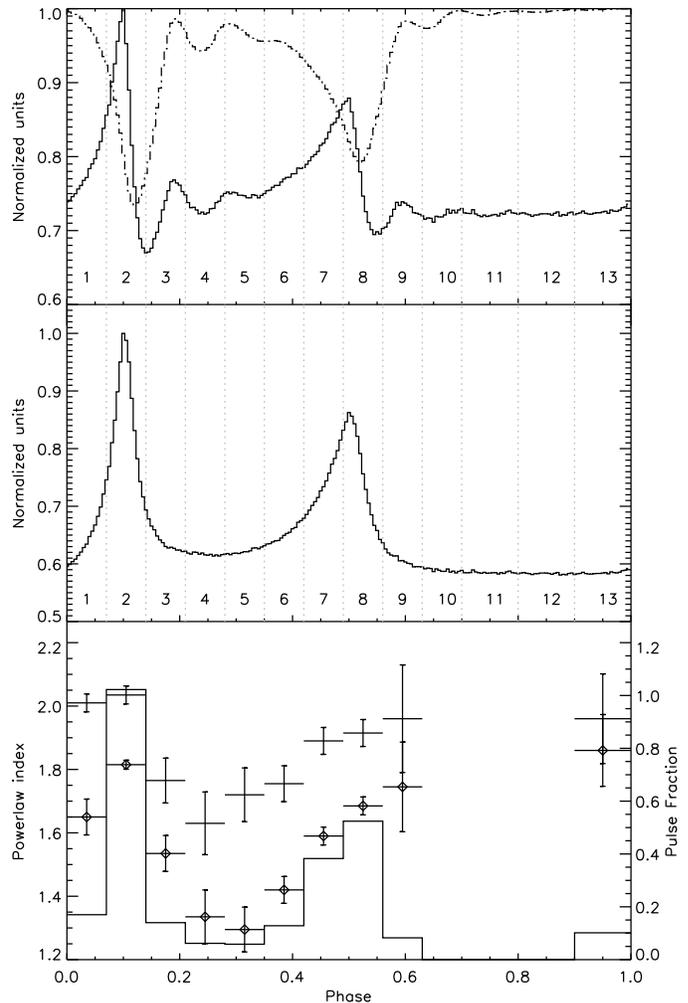}
\end{center}
\caption{Top: The pulse profile (P = 33ms) of module A is shown in solid, the live-time curve is shown dashed, and the boundaries of the extracted phase bins as vertical dotted lines. The minimum of the live-time curve occurs just after the peak and the oscillatory pattern is due to the dead-time of 2.5 ms. Middle: Live-time corrected pulse profile. Bottom: Power-law index averaged between 17\as\ and 50\as\ regions are shown in diamonds for $\Gamma_1$ and crosses $\Gamma_2$. The histogram shows the relative normalization of pulsed emission to the nebular emission ($N_\mathrm{pulse}/N_\mathrm{nebula}$) within each bin.}
\label{pulseprofile}
\end{figure}

\subsection{Crab flaring}\label{flaring}
In addition to the long time scale flux variations observed in the X-ray band, the Crab is known to flare in GeV $\gamma$-rays. Observations with \textit{Agile} \citep{Tavani2011} and \textit{Fermi} \citep{Abdo2011} in the 0.1 -- 1 GeV range have observed flares with a frequency of roughly $\sim1$ a year with typical durations of $\sim10$\,days and a flux increase of about a factor of 5. The radiation is thought to be of synchrotron origin due to the rapidity of the flares and the fast electron cooling time, as opposed to Bremsstrahlung or inverse-Compton emission that have cooling timescales of the order of $\sim10^6-10^7$\,years \citep{Abdo2011}. The emission region is further thought to be Doppler boosted towards the observer \citep{Buehler2012} and causality arguments suggest the flares originate from a very small region, most likely inside the termination shock zone. However, the spatial resolution of current $\gamma$-ray instruments is not sufficient to resolve the inner parts of the nebula. \citet{Weisskopf2013} analyzed \textit{Chandra}, Keck and VLA data during the 2011 April flare to look for such a counterpart, but none of these instruments found conclusive evidence of a change in the Crab emission.

On the 9th of March 2013 \textit{NuSTAR} triggered an observation during one such flaring episode \citep{Mayer2013}. This observation is labeled `CrabToO' in Table \ref{obsid} and had a duration of $\sim$16~ksec and an effective exposure of $\sim$4\,ksec after taking dead time,  SAA passages, and occultation into account.  {\em NuSTAR} caught the Crab at the tail end of flaring and did not detect any spectral variation or flux change beyond the level that can be expected from calibration ($\pm 5$\% in flux and $0.01$ in spectral index change).

\subsection{Phase Resolved Spectroscopy}\label{phaseresolved}

\subsubsection{Pulsar Spectrum}
The Crab pulsar has a 33~ms period and a pulse profile that is double peaked in all energy bands. In the X-rays the primary peak is higher than the secondary, and the off-pulse period spans about 30\% of the phase (see Figure 3, middle panel). The Crab ephemeris is routinely calculated by the Jodrell Bank Observatory \footnote{(http://www.jb.man.ac.uk/pulsar/crab.html)} and we used the closest ephemeris entry for each observation to obtain the pulse profile.

For the phase resolved spectroscopy, all observations listed in Table \ref{obsid} were combined for FPMA, and for FPMB ``Crab34'' was excluded due to proximity of the pulsar to the chip gap. The Ancillary Response Files (ARFs) of the different observations were combined, while the same detector response matrix (RMF) could be used since the pulsar was on detector 0 for all observations. We barycenter corrected event times using the {\tt barycorr} routine that is part of the HEASOFT {\tt FTOOL} library, and corrected the \textit{NuSTAR} clock for thermal drifts. 

The resulting raw (not live-time corrected) pulse profile is shown in the top panel of Figure~\ref{pulseprofile} (solid line). The pulse profile exhibits a distinct oscillatory pattern due to dead time effects associated with the 2.5~msec event readout \citep{Harrison2013}. The decreased probability of detecting an event just after pulse peak results in a pattern of damped oscillations with a period of $\sim$3.4~msec and causes the live-time fraction to vary throughout the pulse period as shown by the dashed curve. To correct for this effect we use the ``PRIOR" column in the event list, which specifies the elapsed time since the prior event. In the absence of events vetoed by the active anti-coincidence shield, this column would accurately reflect the true elapsed live-time, but the standard operating mode does not downlink vetoed events, so that in general adding the PRIOR column would not yield the proper live-time. However, for source count-rates significantly higher than the veto rate the error becomes negligible, and for the Crab  we determine it is only $\sim$0.3\%, by comparing the summed PRIOR column to the live-time reported by the instrument once per second.

We used the live-time curve to adjust the net exposure time of each phase, and for each phase bin we extracted events for both 17\as\ and 50\as\ circular regions centered on the pulsar. The reason for picking two extraction regions is to test the stability of the fitting procedure. The nebular component will change as a function of the extraction region size, while the pulsar component should not, providing a cross check on how well the two components are distinguished. To avoid the statistical pit-falls of subtracting a large component, we decided to fit the two components together, presuming that each phase bin can be decomposed into the un-pulsed nebula, constant throughout all phases, and the pulsar. 

We first investigate what spectral models to use by fitting the three phase bins 10 -- 12 corresponding to the pulse-off interval. We found that a power-law yields an inadequate fit ($\chi^2_\mathrm{red}$ = 1.92 for 1076 degrees of freedom (dof)) and that a broken power-law provides a much better fit ($\chi^2_\mathrm{red}$ = 1.05 for 1074 dof). We also tried out a \texttt{logpar} model and an exponential \texttt{cutoff} model, but both failed to fit the high energy tail of the spectrum and under-predicted the flux. Although we recognize that it is difficult to infer physical meaning from a broken power-law, we chose to continue with this representation because it gives a simple and intuitive understanding of the shape of the spectrum.

We use the best fit broken power-law to investigate the shape of the pulsar component during the peaks in phase bins 2 and 8.  Freezing the broken power-law model to what we derived for phase bins 10 -- 12, we fit the pulsar component with a power-law and a broken power-law respectively. Once again the broken power-law clearly provides a better fit; for phase bin 2 the power-law fit has $\chi^2_\mathrm{red}$=1.19 for 806 dof and broken power-law $\chi^2_\mathrm{red}$=1.00 for 809 dof. On this basis we chose to model with two independent broken power-laws. We fit a total of 26 spectra (13 phase bins for FPMA and FPMB) in XSPEC between 3 -- 78 keV.

We observe degeneracies between the models when we do not constrain the normalization of the pulsar component during phase bins 10 -- 12.  We resolve this by limiting the contribution of the pulsar during these phases to be a small fraction during pulse-on phases. According to \citet{Weisskopf2011}, the ratio of the 0.3 -- 3.8\,keV flux between the primary pulse peak and pulse-off is about a factor of 100, and we applied this restriction to the relative pulsar normalization in the model for these phase bins. We coupled the break energy, BE, between all phase bins for the pulsar and nebula respectively. Table \ref{phaseresolvedtable} shows the best-fit parameters for the pulsar spectrum for both extraction regions in the top half for all phase bins except 10 -- 12. The parameters for the two extraction regions agree for most bins within 1$\sigma$ and all of them within 2$\sigma$. The bottom half lists the best fit parameters for the nebula in the two extraction regions and as anticipated they are different, softening with increasing radius but maintaining the same break energy. Figure~\ref{pulseprofile} bottom panel shows the averaged power-law for each phase bin of the two extraction regions for $\Gamma_1$ and $\Gamma_2$ respectively. We have omitted the parameters for phase bins 10 -- 12 since the pulsar component could not be properly constrained during these periods. The average $\Delta\Gamma = \Gamma_2 - \Gamma_1$ across the phase bins, excluding bins 10 -- 12, is $0.27 \pm 0.09$.


\subsubsection{Spatially Integrated Phase Resolved Spectrum}
We measured the phase-averaged spectrum from a 200\as\ region in Section \ref{phaseaveraged} and found $\Gamma = 2.0963 \pm 0.0004$. At smaller radii we find that the pulsar and the nebula can both be characterized independently by broken power-laws. It may seem strange that the superposition of apparent broken and un-broken power-laws should sum up to a simple power-law. To show how the Crab achieves this, we investigate here the phase-resolved total spectrum (nebula+pulsar) using the same phase bins as defined in the previous section. We use a broken power-law where required by the data, and power-law otherwise and the fit results are listed in Table \ref{80pixfits}.

\begin{figure}
\begin{center}
\includegraphics[width=0.47\textwidth]{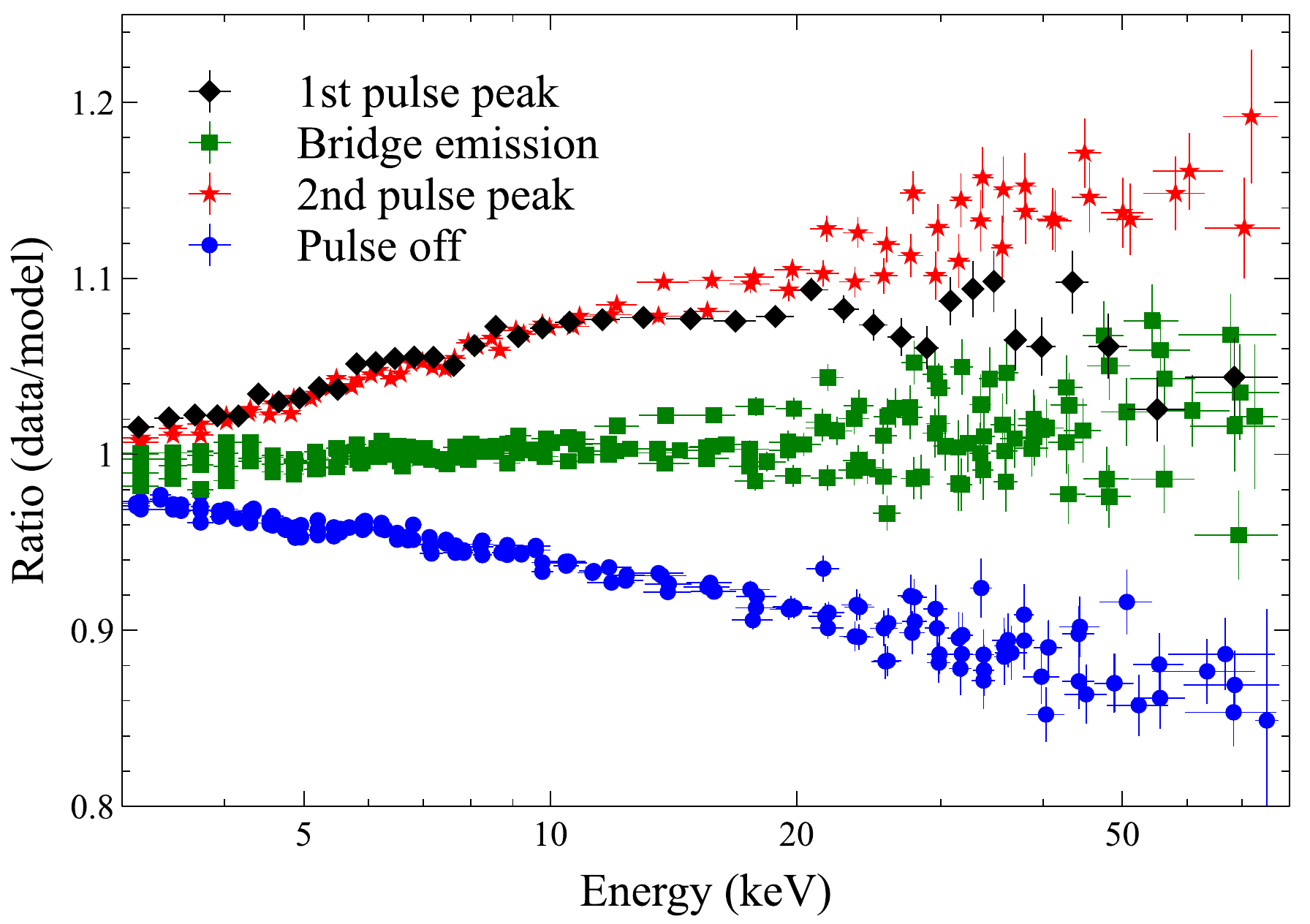}
\end{center}
\caption{Ratio plot of best fit model $\Gamma = 2.1$ of the 13 phase bins extracted from a region of 200\as\ centered on the pulsar.}
\label{pulseprofile80pix}
\end{figure}

Figure \ref{pulseprofile80pix} shows the ratio of the data to the phase-average power-law fit. The curves have been scaled for clarity to show how the Crab roughly decomposes into four components; 1st pulse, bridge emission, 2nd pulse, pulse-off (nebula). Although both pulse peaks have spectra that steepen ($\Gamma_2 \sim 2.047 - 2.099$), the spectra remain harder than the phase-averaged value. The bridge emission is best approximated by a power-law with an index close to the phase-averaged value ($\Gamma = 2.086 \pm 0.002$), while the off-pulse nebula emission is significantly softer and mildly broken ($\Gamma_1 = 2.123 \pm 0.003$, $\Gamma_2 = 2.134 (-0.004/+0.01)$). The sum of nebula and pulsed emission thus approximately conspire to mimic the phase-averaged power-law and explain why as a whole we observe a power-law spectrum.

\begin{figure}
\begin{center}
\includegraphics[width=0.47\textwidth]{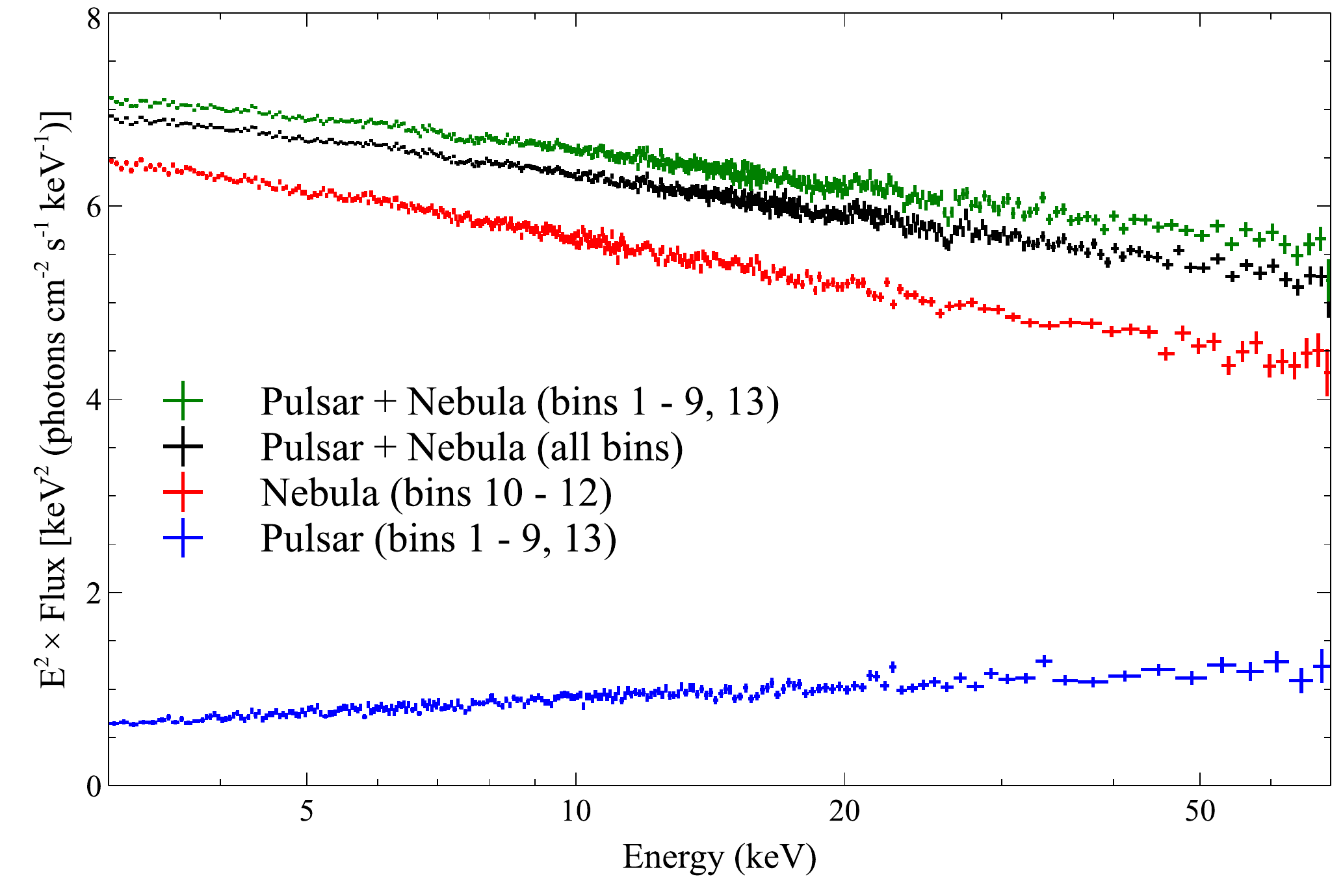}
\end{center}
\caption{Spectral energy distribution of the pulsar + nebula over all phase bins, nebula (phase bins 10 -- 12), pulsar + nebula (phase bins 1 -- 9, 13), and pulsar (phase bins 1 -- 9, 13) alone with nebula subtracted.}
\label{psr_p_pwn}
\end{figure}

Finally in Figure \ref{psr_p_pwn} we show the spectral energy distribution for the (1) phase averaged pulsar + nebula, (2) nebula (phase bins 10 -- 12), (3) pulsar + nebula (phase bins 1 -- 9, 13), (4) and pulsar (phase bins 1 -- 9, 13) with nebula subtracted. The fit parameters for (1) is $\Gamma = 2.0963 \pm 0.0004$, (2) $\Gamma = 2.0865 \pm 0.0004$, (3) nebula alone as recorded in Table \ref{80pixfits}, (4) and for the pulsar alone $\Gamma_1 = 1.72 \pm 0.02$, $\Gamma_2 = 1.72 \pm 0.02$ and BE = $10.5 \pm 1.6$.

\begin{table*}
\caption{Phase resolved fits to extraction regions 17\as\ and 50\as}
\centering
\begin{tabular}{l|ccc|ccc}
\hline
\multicolumn{1}{l}{} & \multicolumn{3}{c}{\texttt{Tbabs(bkn+bkn)}} & 
\multicolumn{3}{c}{\texttt{Tbabs(bkn+bkn)}} \\
\multicolumn{1}{l}{} & \multicolumn{3}{c}{17\as: $\chi^2_\mathrm{red}$=1.005 (for 15884 degrees of freedom)} & 
\multicolumn{3}{c}{50\as: $\chi^2_\mathrm{red}$=1.018 (for 23752 degrees of freedom)} \\
\multicolumn{1}{l}{} & \multicolumn{3}{c}{Pulsar} & \multicolumn{3}{c}{Pulsar} \\
\multicolumn{1}{l}{} & \multicolumn{3}{c}{\texttt{bknpower}} & \multicolumn{3}{c}{\texttt{bknpower}} \\
\hline
Phase & $\Gamma_1$ & $\Gamma_2$ & BE\footnote{Break energy (BE) component coupled for all phase bins.} (keV)  & $\Gamma_1$ & $\Gamma_2$ & BE$^\mathrm{a}$ (keV) \\
\hline
1  (0-0.07) &1.64$\pm$0.04 & 2.01$\pm$0.02 & 11.7$\pm$0.6 & 1.66$\pm$0.04 & 2.01$\pm$0.02 & 13.1$\pm$0.4 \\
2  (0.07-0.14) &1.80$\pm$0.01 & 2.02$\pm$0.02 &.               & 1.83$\pm$0.01 & 2.05$\pm$0.02 &. \\
3  (0.14-0.21) &1.52$\pm$0.04 & 1.76$\pm$0.05 &.               & 1.55$\pm$0.04 & 1.77$\pm$0.05 &. \\
4  (0.21-0.28) &1.39$\pm$0.06 & 1.62$\pm$0.07 &.               & 1.28$\pm$0.06 & 1.64$\pm$0.07 &.  \\
5  (0.28-0.35) &1.30$\pm$0.05 & 1.70$\pm$0.06 &.               & 1.29$\pm$0.05 & 1.74$\pm$0.06 &.  \\
6  (0.35-0.42) &1.42$\pm$0.03 & 1.75$\pm$0.04 &.               & 1.42$\pm$0.03 & 1.76$\pm$0.04 &. \\
7  (0.42-0.49) &1.57$\pm$0.02 & 1.93$\pm$0.03 &.               & 1.61$\pm$0.02 & 1.85$\pm$0.03 &. \\
8  (0.49-0.56) &1.66$\pm$0.02 & 1.92$\pm$0.03 &.               & 1.71$\pm$0.02 & 1.91$\pm$0.03 &.  \\
9  (0.56-0.63) &1.66$\pm$0.08 & 1.97$\pm$0.12 &.               & 1.83$\pm$0.08 & 1.95$\pm$0.12 &.  \\
13 (0.9-1.0) & 1.76$^{+0.2}_{-0.08}$ & $1.81\pm$0.10 & .            & 1.96$\pm$0.08 & 2.11$\pm$0.10 &.  \\
\hline
\multicolumn{7}{c}{}\\
\hline
\multicolumn{1}{l}{} & \multicolumn{3}{c}{Nebula} & \multicolumn{3}{c}{Nebula} \\
\multicolumn{1}{l}{} & \multicolumn{3}{c}{\texttt{bknpower}} & \multicolumn{3}{c}{\texttt{bknpower}} \\
\hline
Phase & $\Gamma_1$ & $\Gamma_2$ & BE (keV)  & $\Gamma_1$ & $\Gamma_2$ & BE (keV) \\
\hline
10-12\footnote{Nebula component only.} (0.63-0.9) & 1.92$\pm$0.01 & 2.00$\pm$0.01 & 8.3$\pm$0.5 & 1.99$\pm$0.01 & 2.09$\pm$0.01 & 8.3$\pm$0.2\\
\hline
\end{tabular}
\label{phaseresolvedtable}
\end{table*}

\begin{table}
\caption{Phase resolved Pulsar+Nebula fits to 200\as\ extraction region.}
\centering
\begin{tabular}{l|lll}
\hline
Phase & $\Gamma_1$ & $\Gamma_2^\mathrm{a}$ & BE\footnote{ If $\Gamma_2$ and break energy (BE) is not given the best fit was a power-law.} (keV)\\
\hline
1  (0-0.07) & 2.090 $\pm 0.003$ & 2.110 $\pm 0.006$ & 11 $\pm 3$ \\
2  (0.07-0.14) & 2.041 $\pm 0.003$ & 2.099$^{+0.006}_{-0.01}$ & 12.3 $\pm 1.2$ \\
3  (0.14-0.21) & 2.081 $\pm 0.002$ & - & - \\
4  (0.21-0.28) & 2.090 $\pm 0.002$ & - & - \\
5  (0.28-0.35) & 2.086 $\pm 0.002$ & - & - \\
6  (0.35-0.42) & 2.064 $\pm 0.002$ & - & - \\
7  (0.42-0.49) & 2.030 $\pm 0.002$ & 2.047$^{+0.006}_{-0.01}$ & 13.7 $\pm 3$ \\ 
8  (0.49-0.56) & 2.034 $\pm 0.003$ &  2.056 $\pm 0.005$ & 10 $\pm 1.5$ \\ 
9  (0.56-0.63) & 2.120 $\pm 0.002$ & - & - \\
10 (0.63-0.7) & 2.120 $\pm 0.004$ & 2.138$^{+0.004}_{-0.01}$ & 10 $\pm 4$ \\
11 (0.7-0.8) & 2.123 $\pm 0.003$ & 2.134$^{+0.004}_{-0.01}$ & 10 $\pm 4$ \\
12 (0.8-0.9) & 2.126 $\pm 0.002$ & - & - \\
13 (0.9-1.0) & 2.124 $\pm 0.002$ & - & - \\
\hline
\end{tabular}
\label{80pixfits}
\end{table}

\subsection{Spatially Resolved Spectroscopy of the Nebula}\label{spatialspectral}

\begin{figure}
\includegraphics[width=0.47\textwidth]{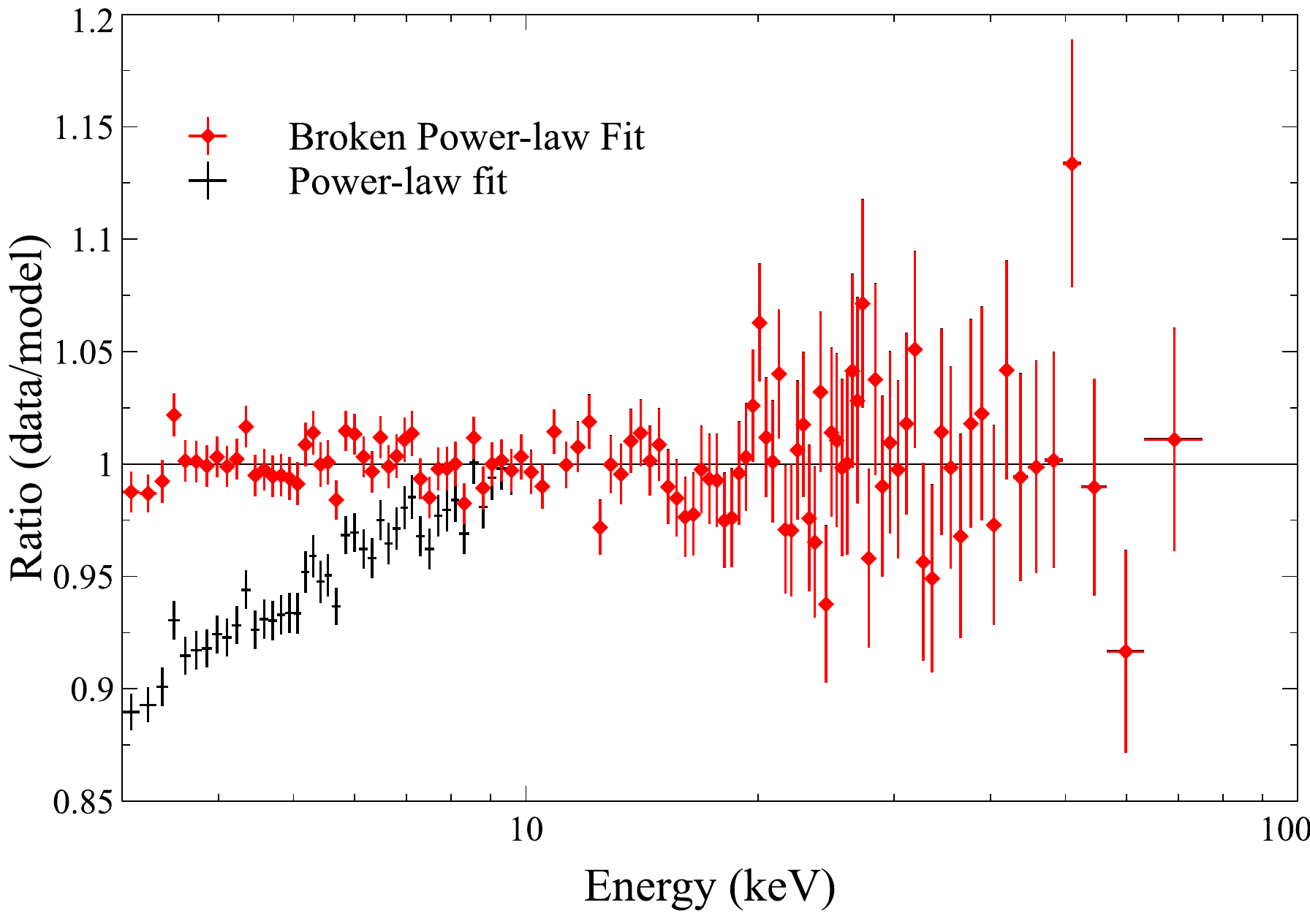}
\caption{Spectrum for off-pulse bins 10 -- 12 extracted from map location (Ra,dec)=(0,0)  of FPMA, which coincides with the pulsar location. The black and red data set is the same, but the models different. The red curve shows the best fit broken power-law, which had $\Gamma_1=1.91 \pm 0.01$, $\Gamma_2=2.03\pm0.01$, E$_{break}=8.7\pm0.9$\,keV. The black curve is a power-law with index $\Gamma = \Gamma_2$ and the normalization scaled such that the high energy part is the same.}
\label{bknfit}
\end{figure}

\begin{figure*}
\includegraphics[width=9cm, viewport=0 200 600 725]{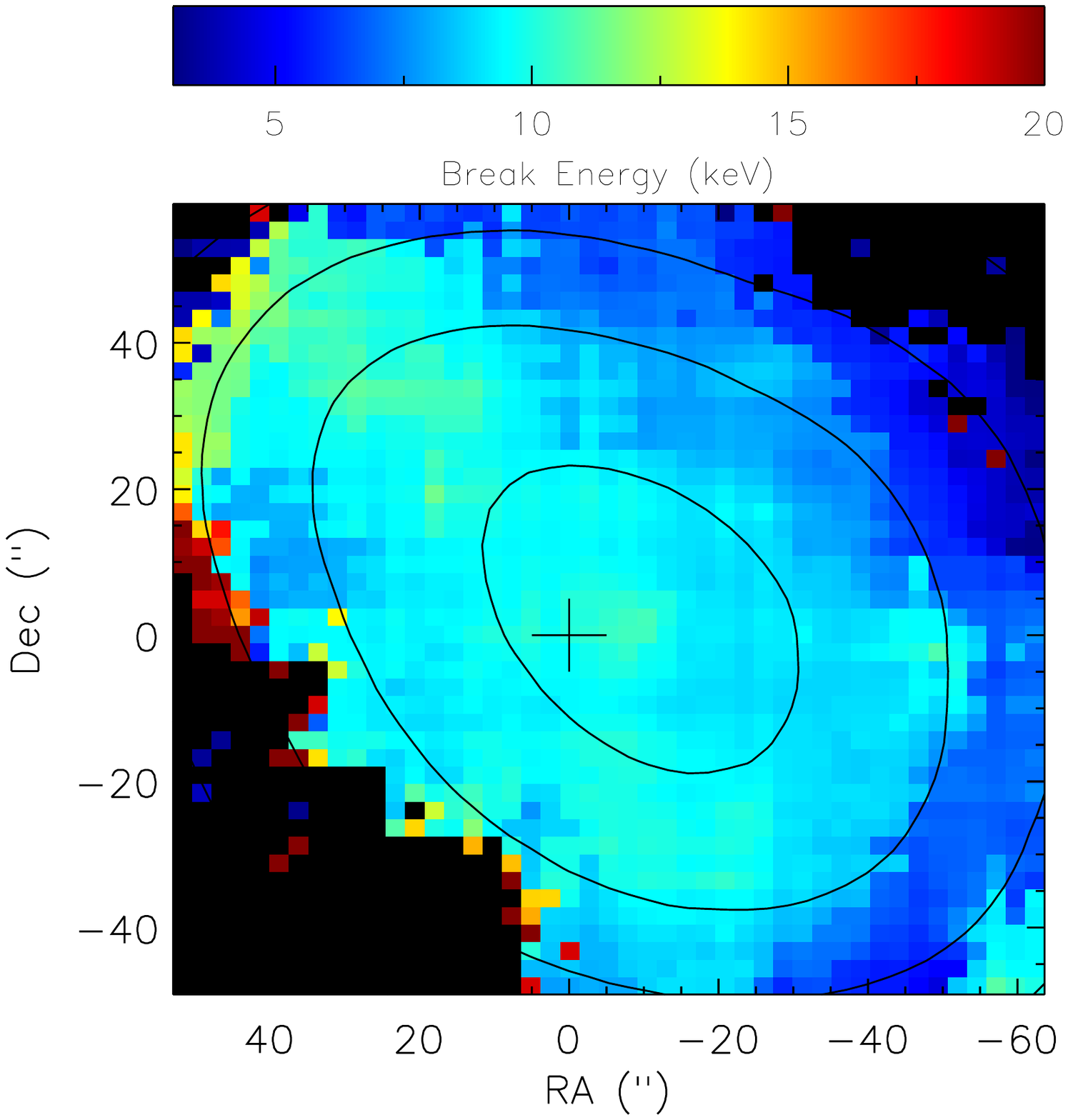}
\includegraphics[width=9cm, viewport=0 200 600 725]{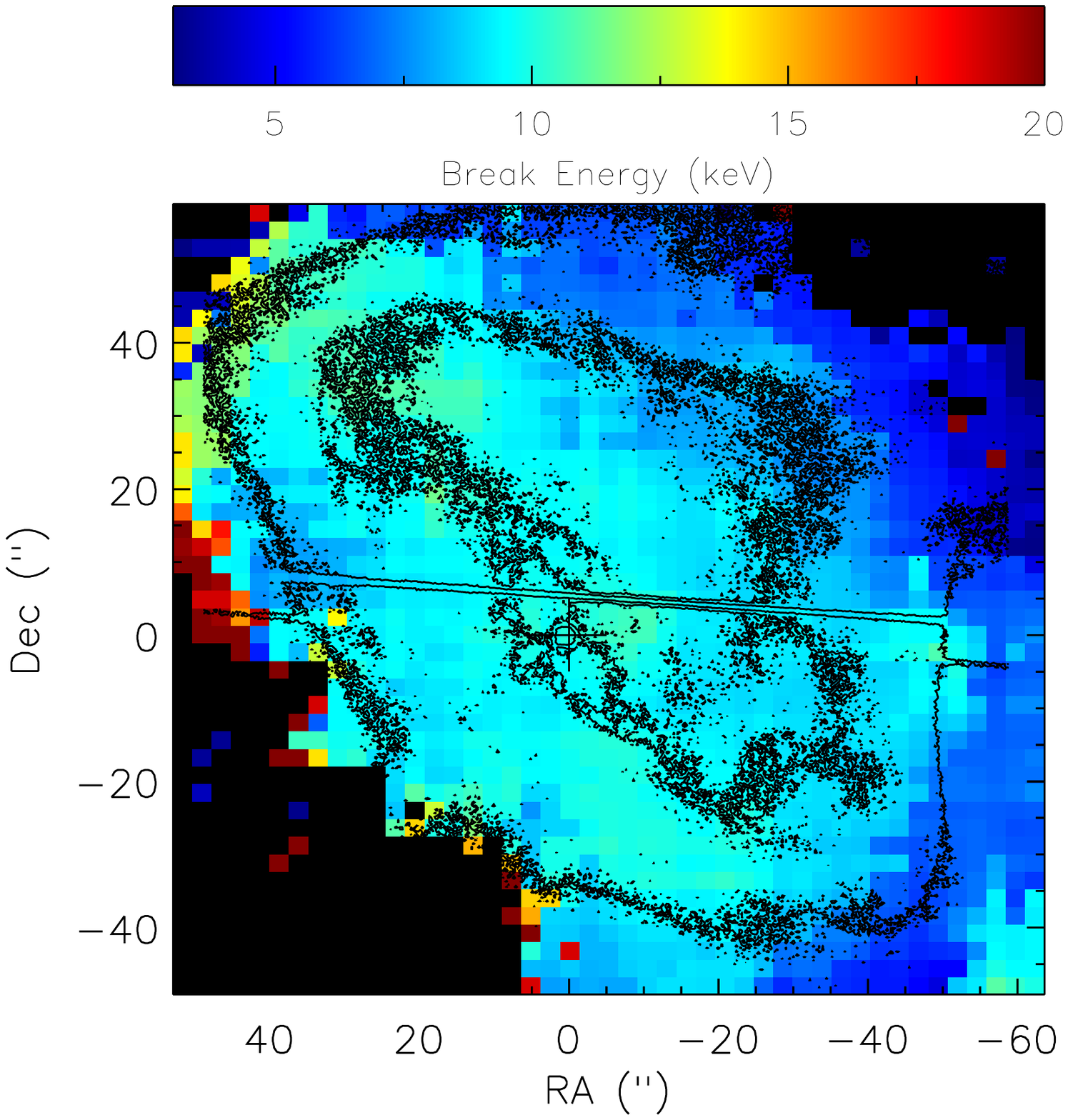}
\includegraphics[width=9cm, viewport=0 200 600 725]{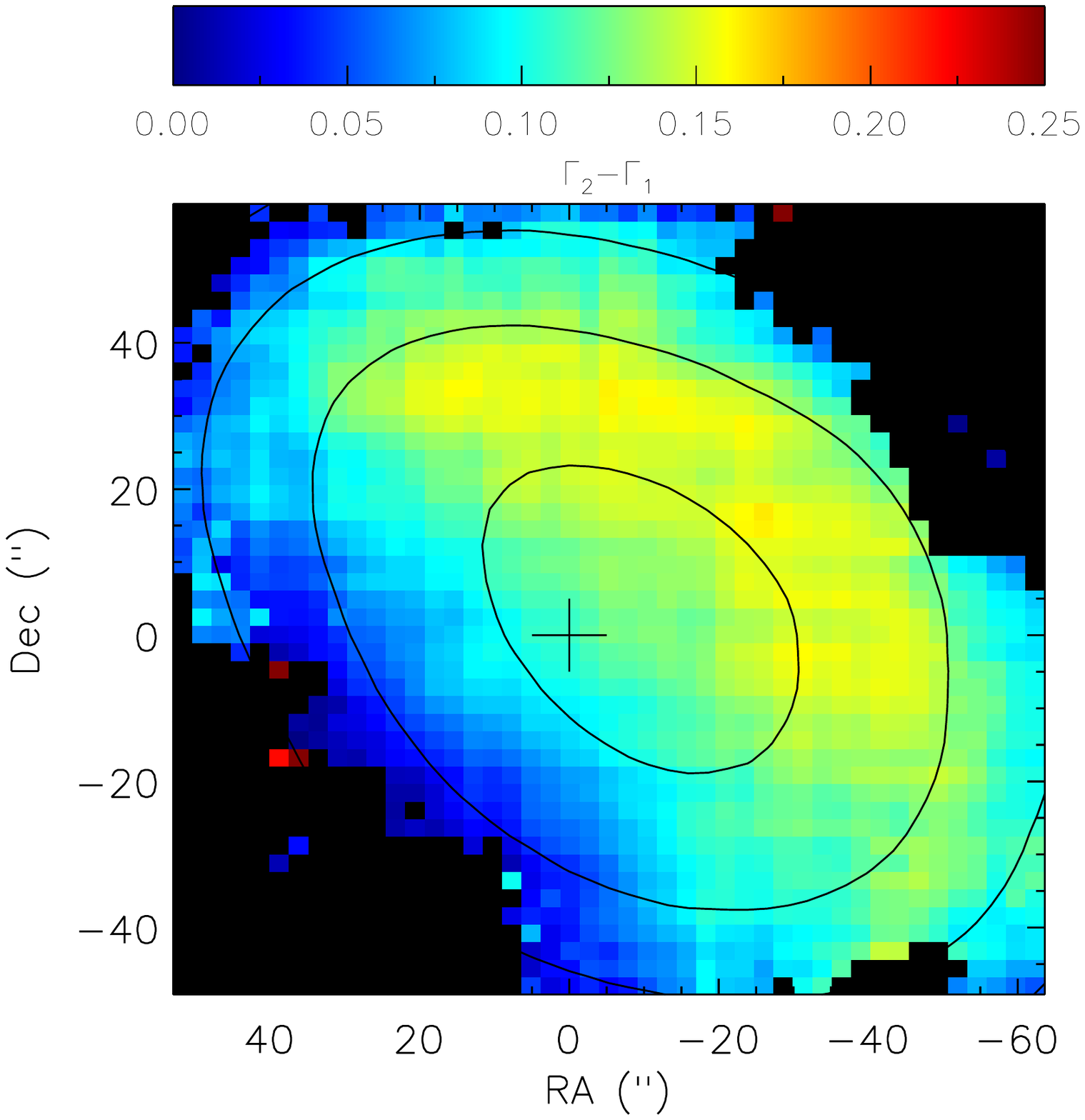}
\includegraphics[width=9cm, viewport=0 200 600 725]{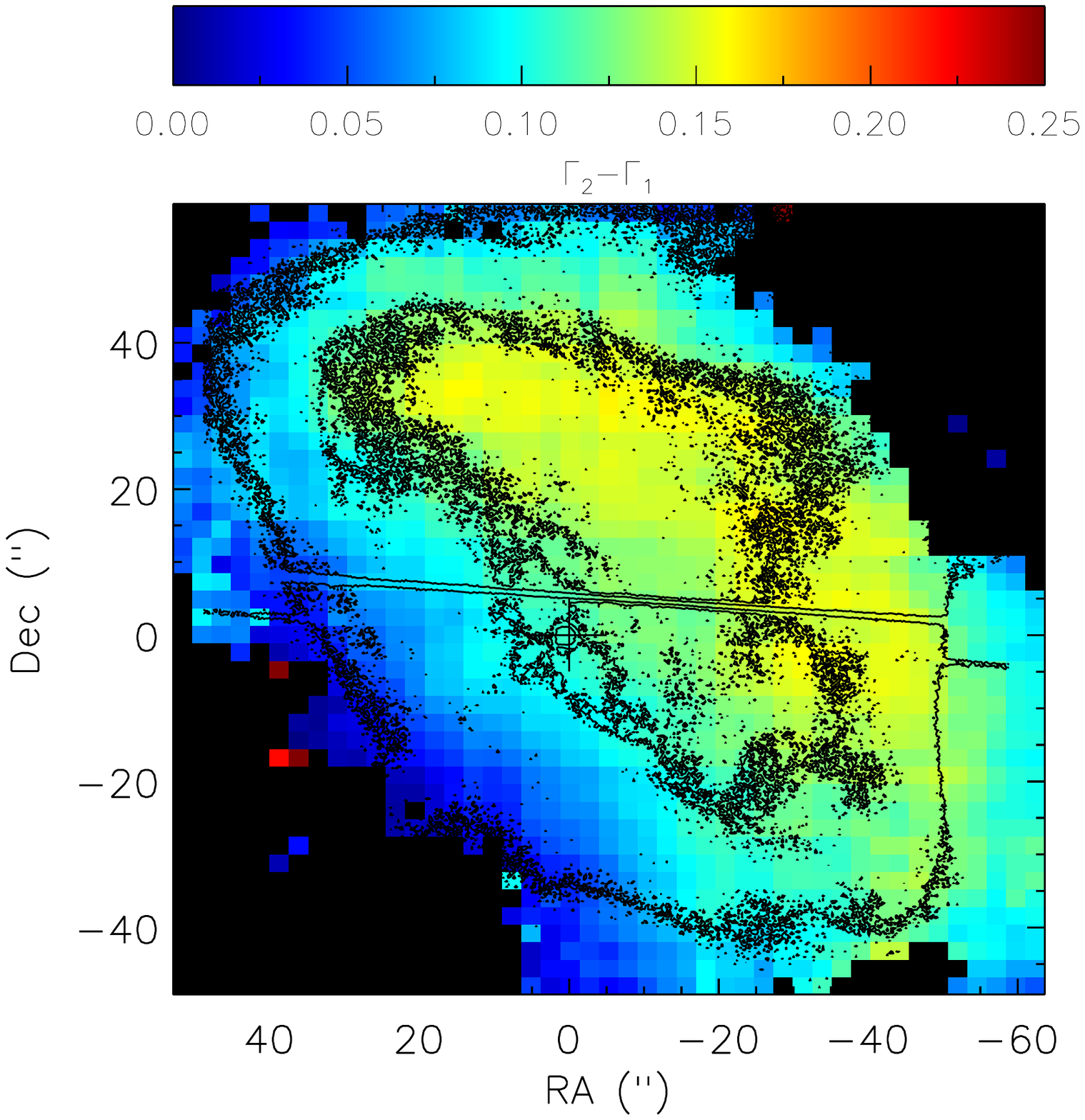}
\caption{Plots showing the break energy and $\Delta\Gamma$ map of $\Gamma_2$-$\Gamma_1$ for FPMB during pulse off. Left panel contours are the \textit{\textit{NuSTAR}} intensity levels and the cross marks the pulsar location. Right panel contours are from \textit{\textit{Chandra}}.}
\label{contour1}
\end{figure*}

\begin{figure*}
\includegraphics[width=9cm, viewport=0 200 600 725]{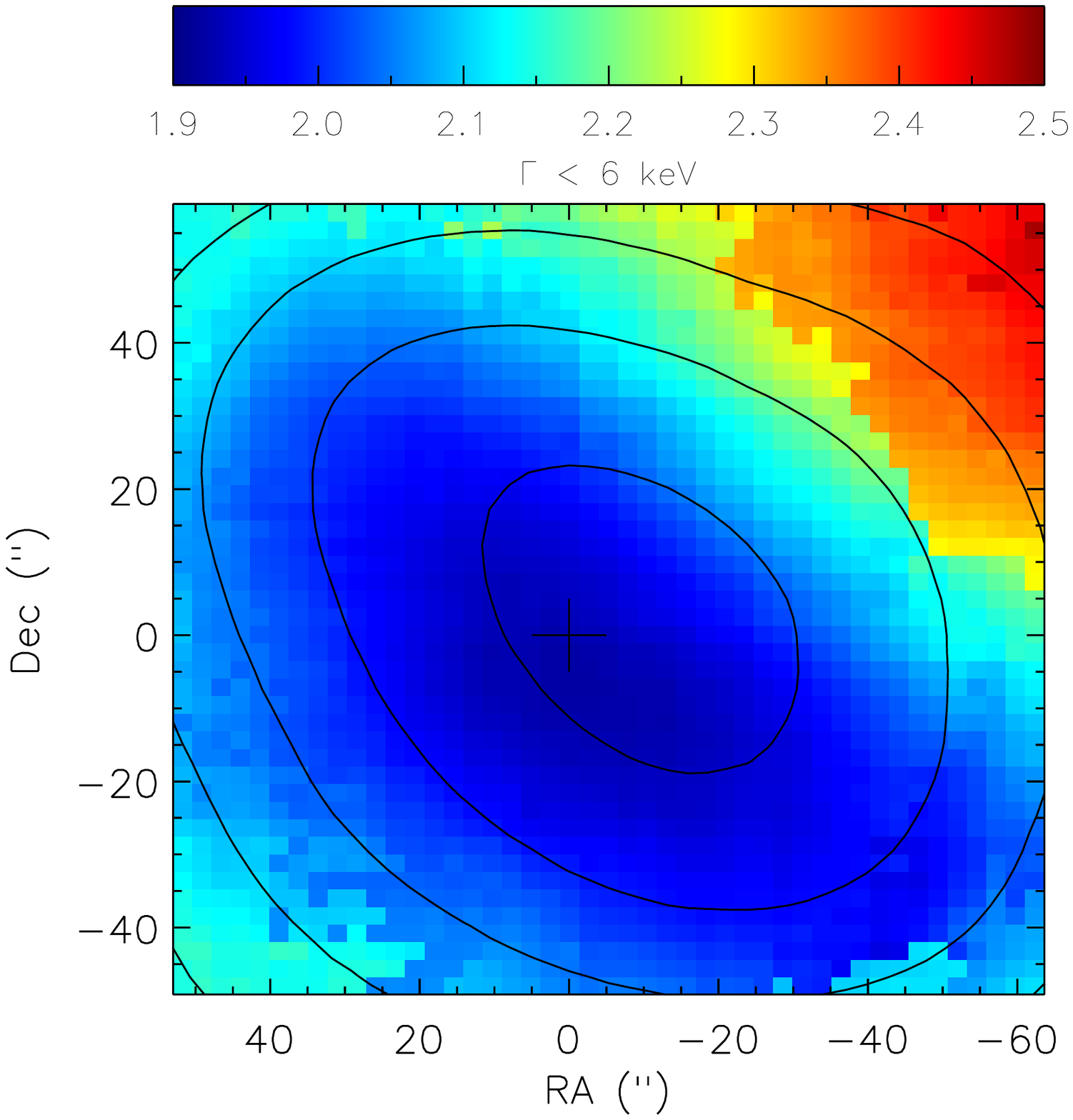}
\includegraphics[width=9cm, viewport=0 200 600 725]{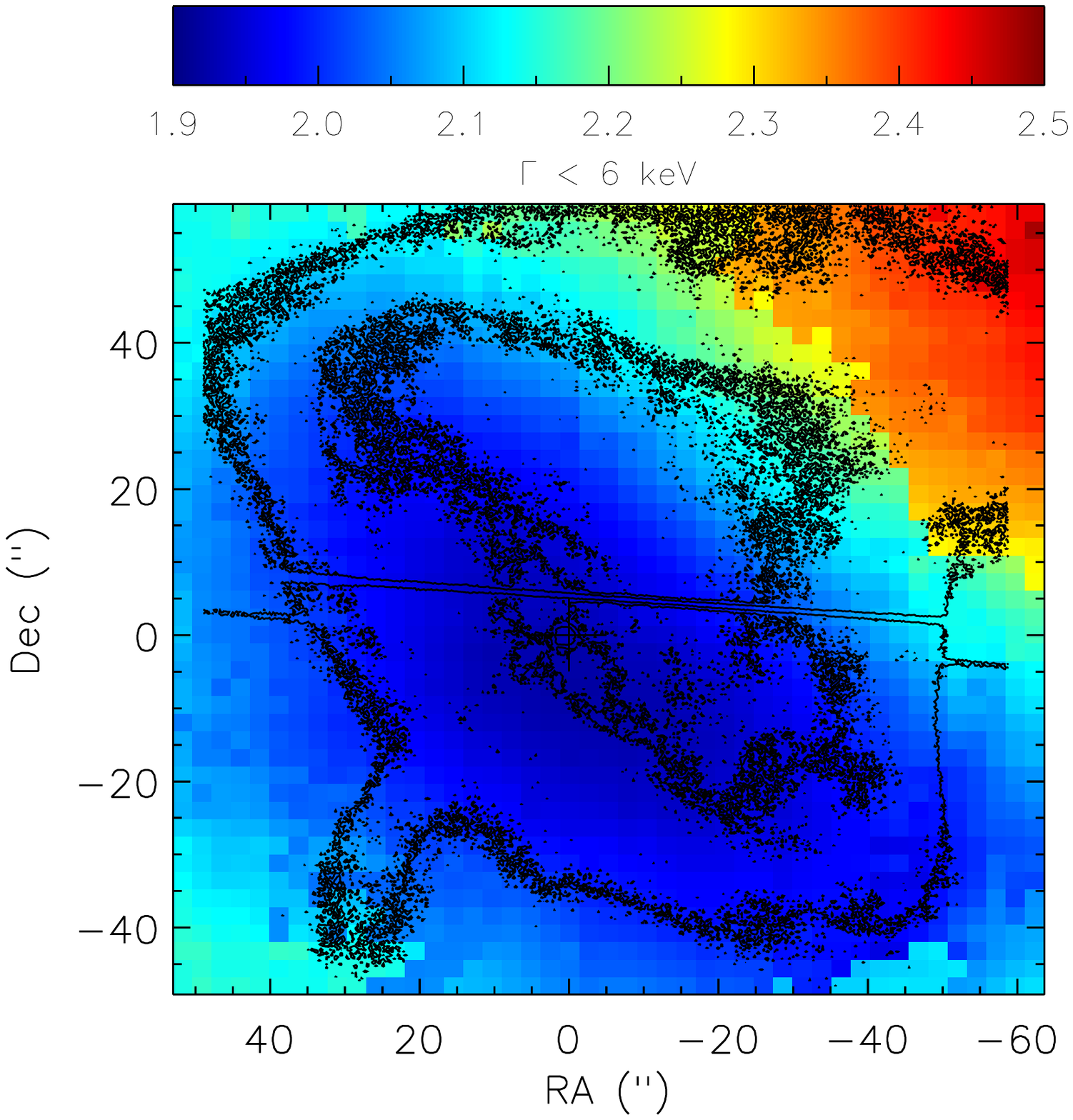}
\includegraphics[width=9cm, viewport=0 200 600 725]{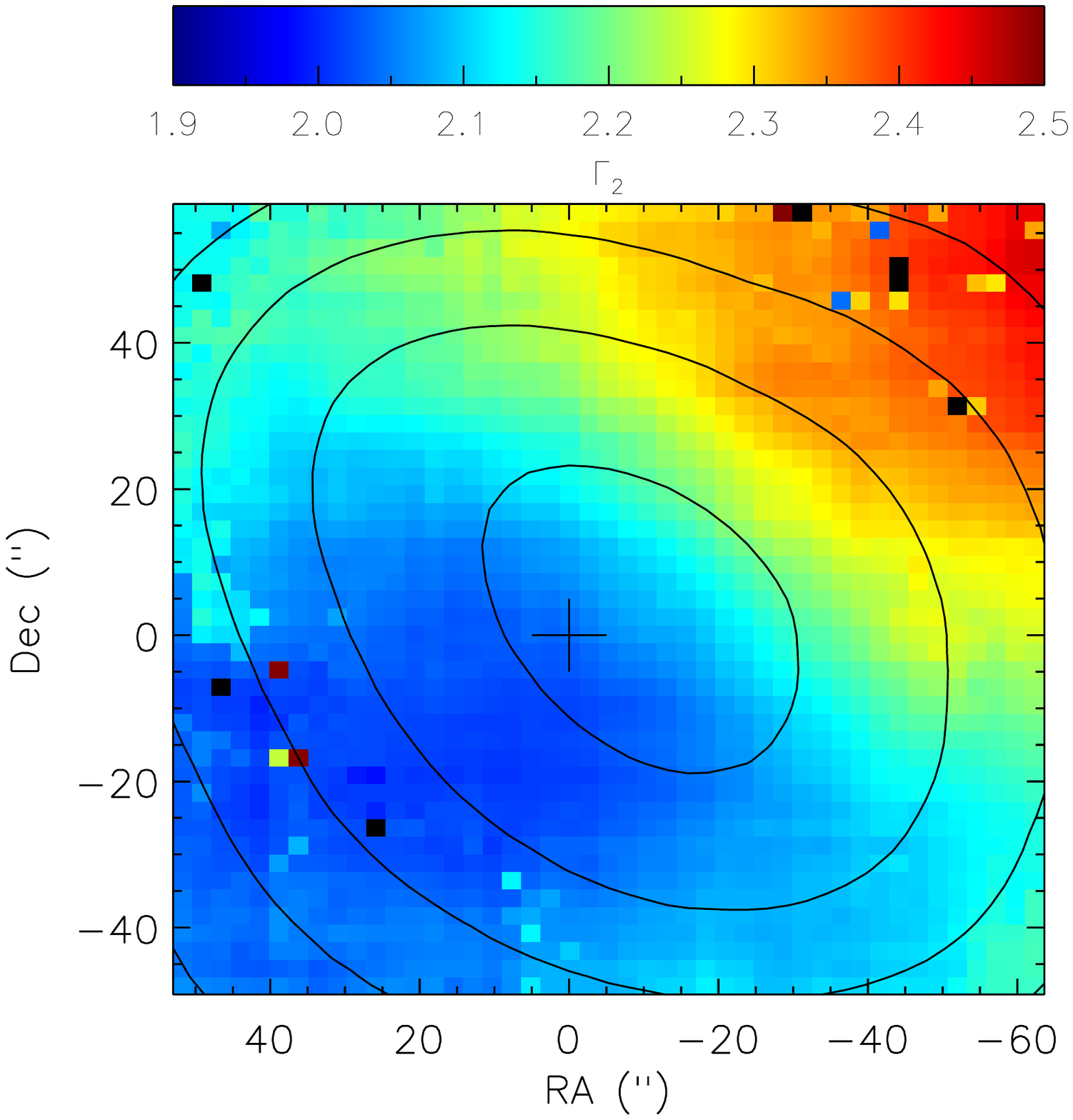}
\includegraphics[width=9cm, viewport=0 200 600 725]{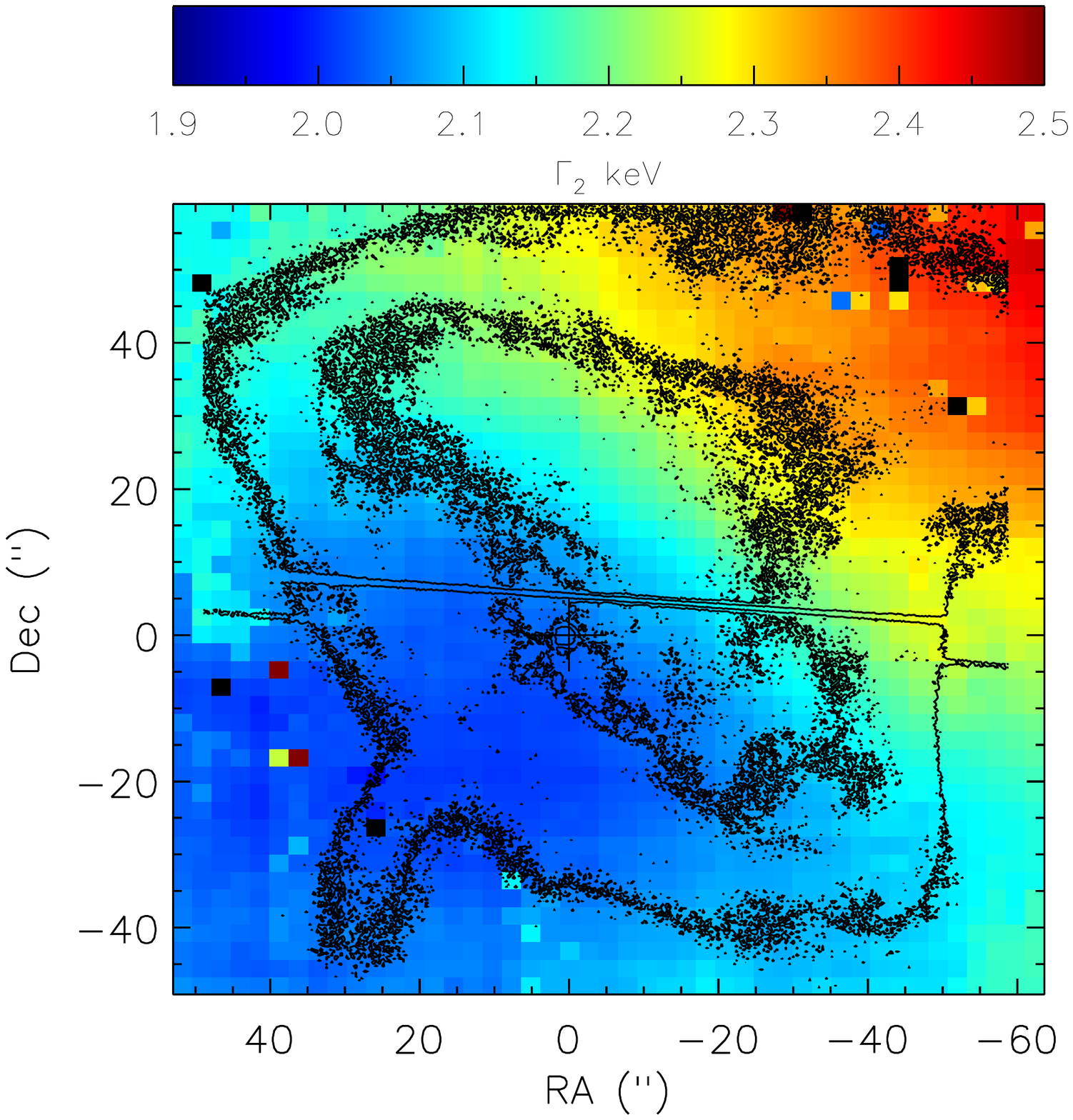}
\caption{Plots power-law map of $\Gamma_{<6keV}$ and $\Gamma_2$ for FPMB during pulse off. Left panel contours are the \textit{\textit{NuSTAR}} intensity levels and the cross marks the pulsar location. Right panel contours are from \textit{\textit{Chandra}}.}
\label{contour2}
\end{figure*}

\begin{figure}
\includegraphics[width=9cm, viewport=0 200 600 725]{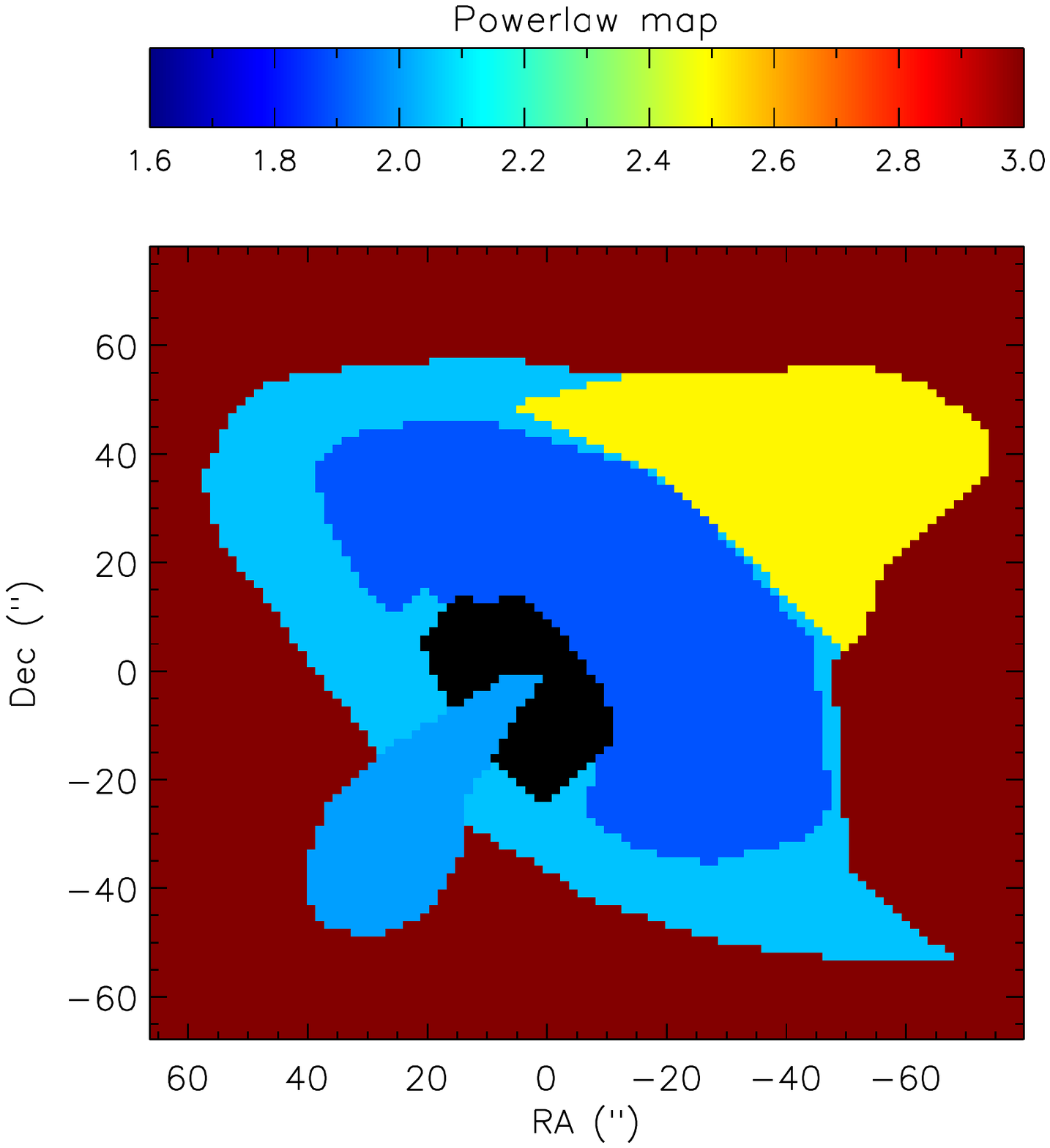}
\includegraphics[width=9cm, viewport=0 200 600 725]{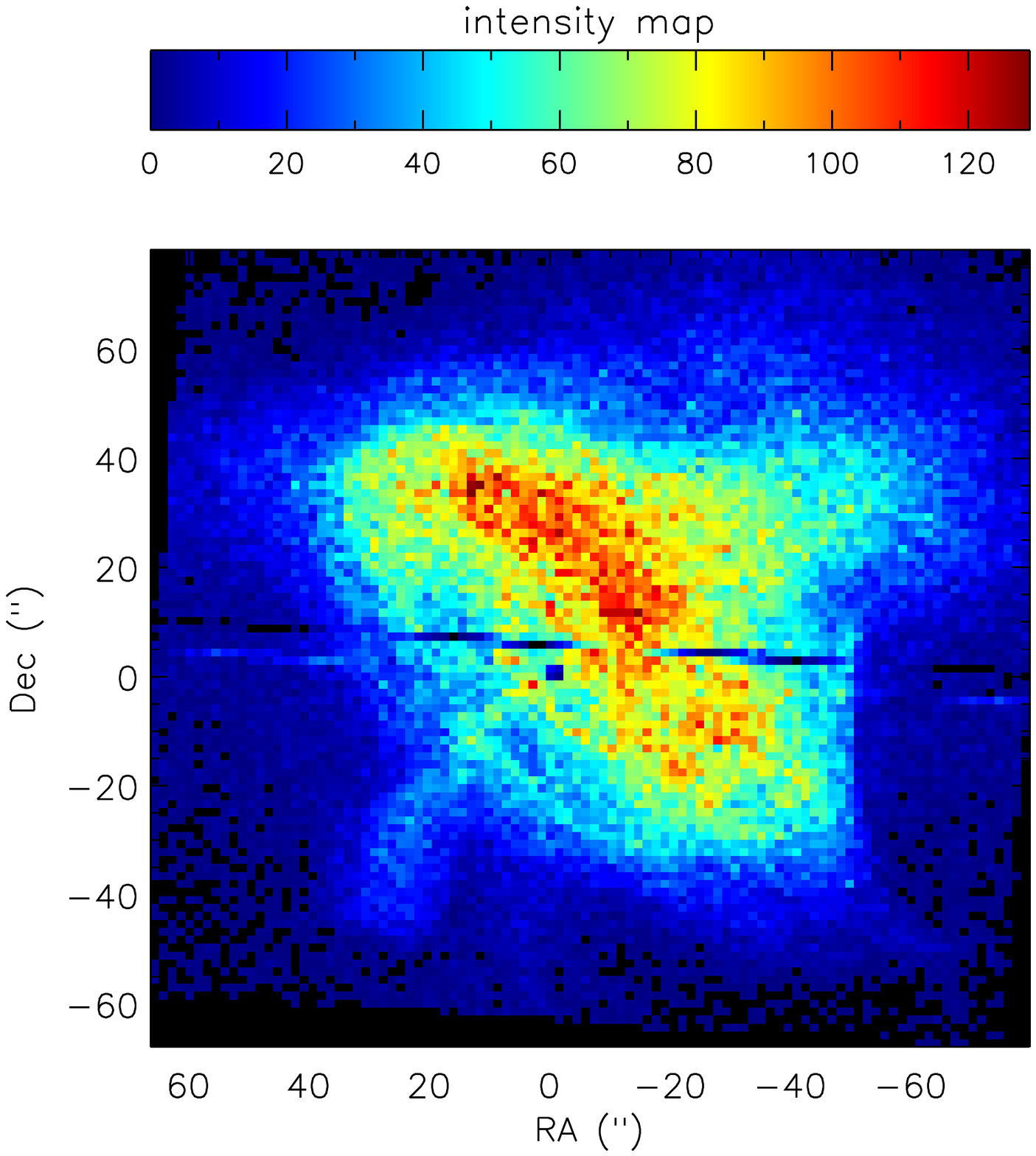}
\caption{Top: Input power-law index map for the simulations. For more details see description in text. Bottom: \textit{Chandra} intensity map used to set the 2 -- 10\,keV flux level for the simulations.}
\label{contour3}
\end{figure}

\begin{figure*}
\includegraphics[width=9cm, viewport=0 200 600 725]{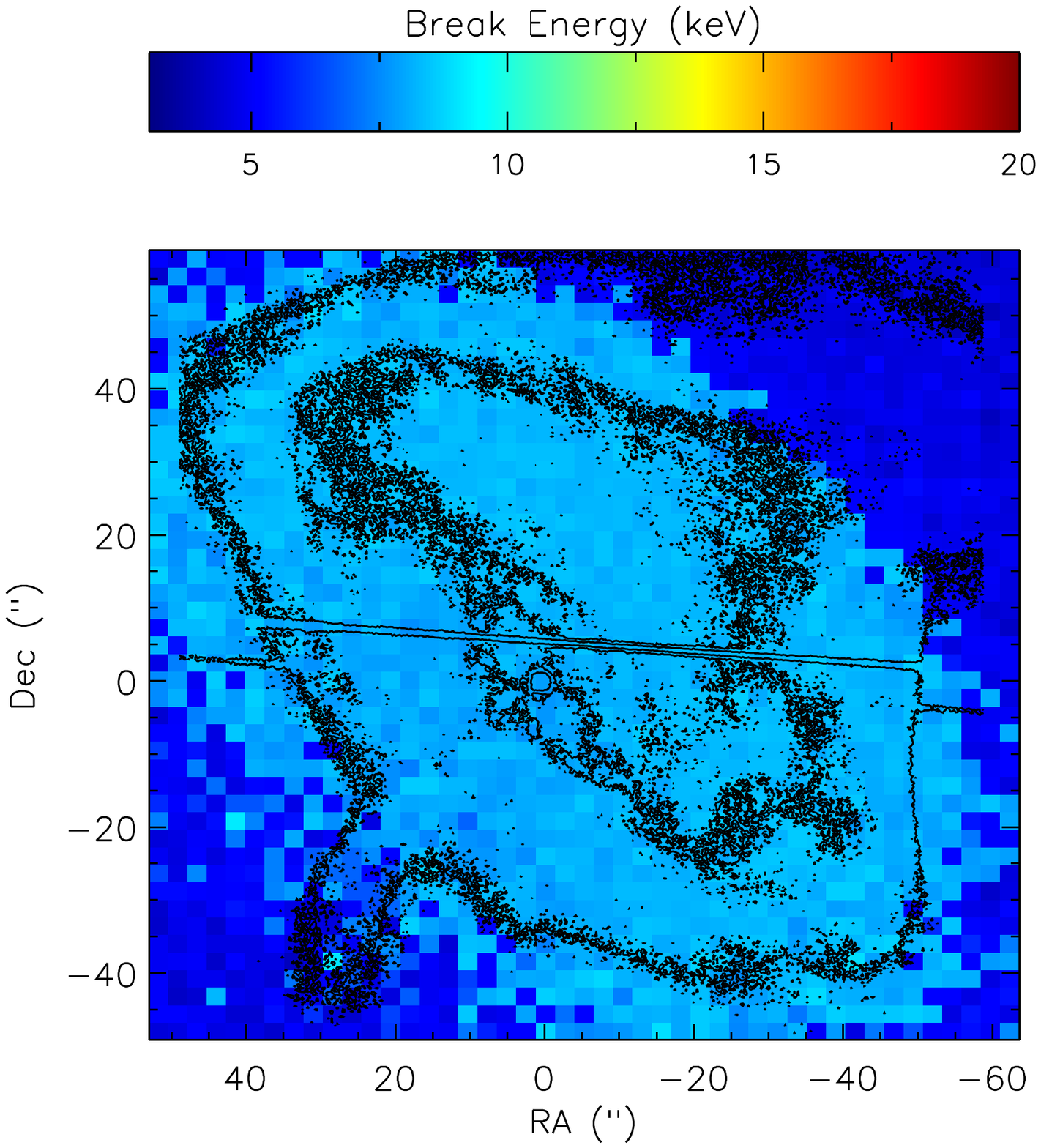}
\includegraphics[width=9cm, viewport=0 200 600 725]{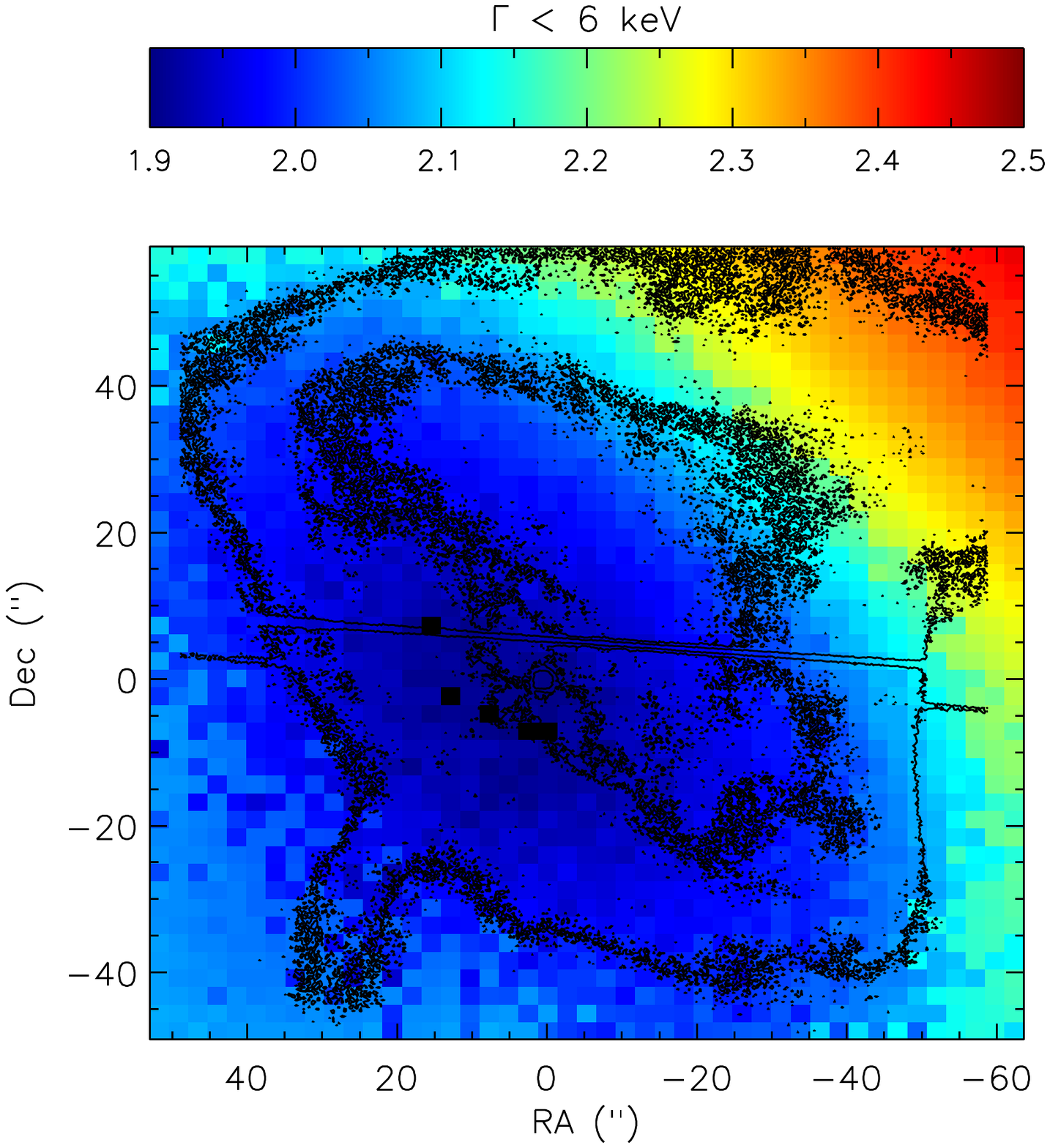}
\includegraphics[width=9cm, viewport=0 200 600 725]{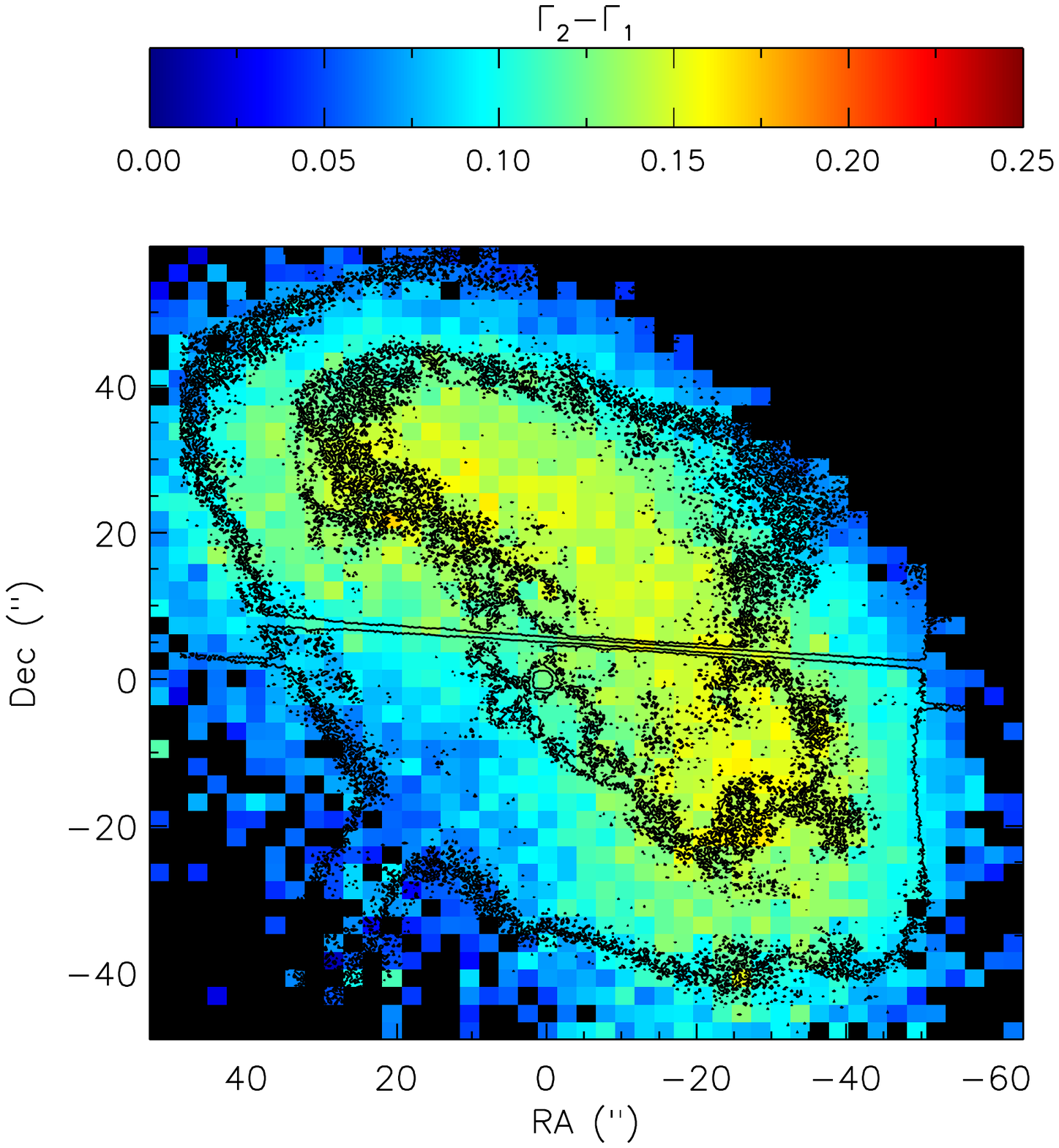}
\includegraphics[width=9cm, viewport=0 200 600 725]{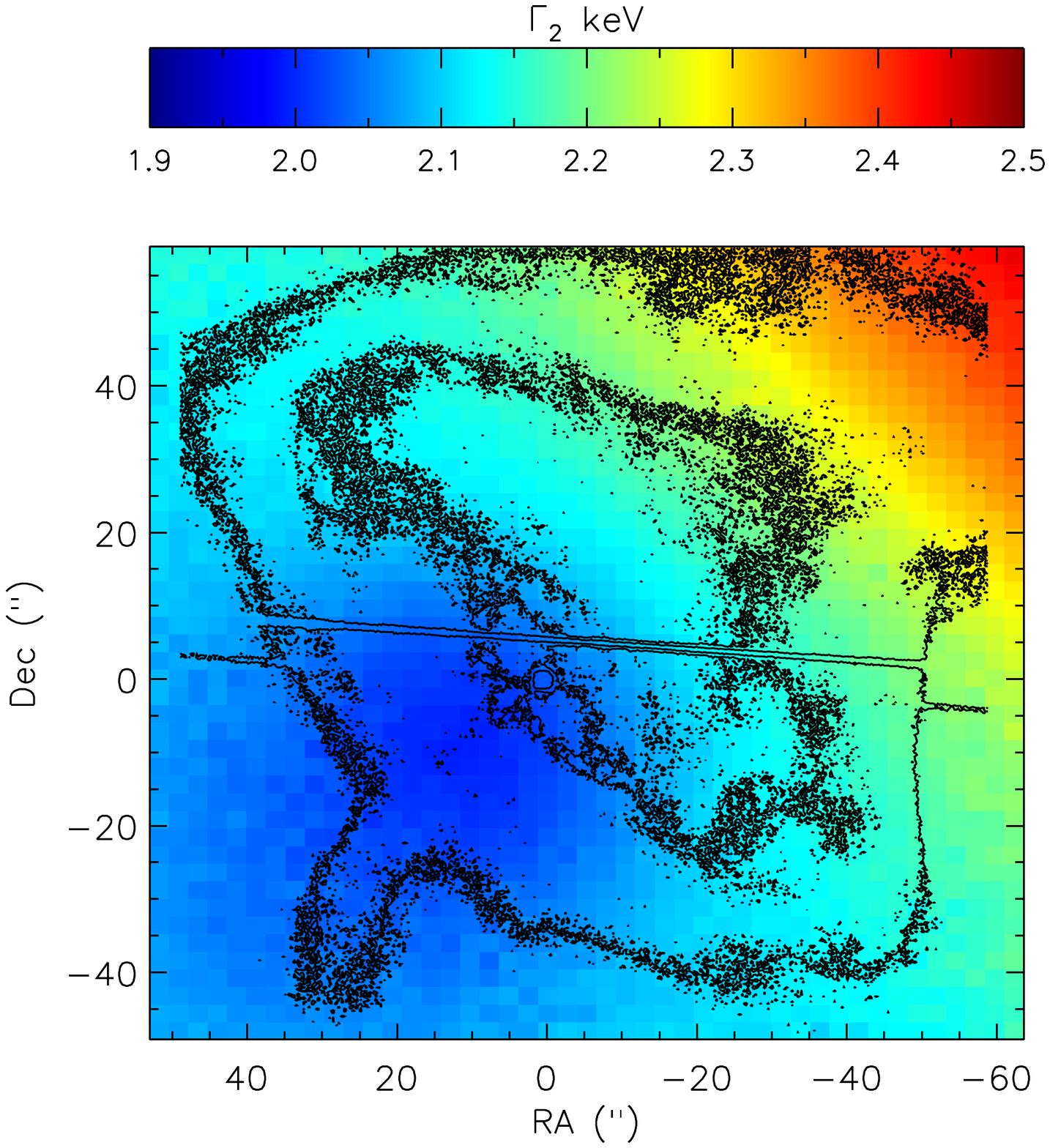}
\caption{Simulated model. Left panel: Break energy and $\Delta\Gamma$ map of $\Gamma_2$-$\Gamma_1$. Right panel: power-law map of $\Gamma_{<6keV}$ and $\Gamma_2$. Contours are from \textit{\textit{Chandra}}}
\label{contour4}
\end{figure*}

From the phase-averaged spectra shown in Figure \ref{cumulative}, it is evident the Crab spectral index changes quite dramatically across the face of the remnant. This is in part due to the harder pulsar spectrum mixing with the softer nebula.  When restricting only to phase bins 10 -- 12 the change is less dramatic, going from $\Gamma_1/\Gamma_2 = 1.92/2.00$ at 17\as, 1.99/2.09 at 50\as\ (Table \ref{phaseresolvedtable}) to 2.12/2.14 at 200\as\ (Table \ref{80pixfits}). Softening of the spectra with increasing radius is predicted in theory and has been observed in G21.5-0.9 \citep{Nynka2014} and 3C 58 \citep{Slane2004} as well in the Crab at lower energies by \textit{Chandra} \citep{Weisskopf2000} and \textit{XMM-Newton} \citep{Kirsch2006}. Due to mixing of spectra at different radial locations by the PSF in \textit{NuSTAR}, it is not straight forward to measure the true (unmixed) index as a function of radius, and a more careful analysis is required to confirm and quantify the effect.

The theoretical predictions apply to the nebula, and we chose to analyze the spatial variation during the phase bins 10 -- 12 when the pulsar is off. We slide a box of 24.5\as$\times$ 24.5\as\ solid angle across the inner 100\as\ to obtain a high S/N spectra, and the box is incremented along RA and Dec by a solid angle of 2.45\as, the size of a {\em NuSTAR} projected sky pixel. At each step we calculate the responses for the box center and fit a broken power-law spectrum. We scale the background from the simulated master background. We performed fitting using XSPEC and Cash statistics.

In general a broken power-law adequately describes the data.  However, in some cases the fitting found either $\Gamma_1=\Gamma_2$, or a break energy less than 5\,keV.  In these cases we reverted to a single power-law model.  At the edges where the torus transitions into the extended nebula, the spectra become complex and can no longer be represented by a broken power-law either. Some of them show negative breaks and for the edge regions we therefore characterized the spectrum with a simple single power-law setting $\Gamma = \Gamma_2$.

In the following we switch to a coordinate system relative to the pulsar location ($\Delta$Ra,$\Delta$Dec)=(0,0). To illustrate the data quality and the level of deviation from a power-law in the center of the remnant, Figure \ref{bknfit} shows the ratio of the spectrum extracted from FPMA for one grid point (24.5\as$\times$ 24.5\as) with the center at ($\Delta$Ra,$\Delta$Dec)=(0,0) to the model. The red curve shows the best fit broken power-law, for which we find $\Gamma_1=1.91 \pm 0.01$, $\Gamma_2=2.03\pm0.01$, and E$_{break}=8.7\pm0.9$\,keV. The black curve shows a single power-law where we have fixed the index $\Gamma = \Gamma_2 = 2.03\pm0.01$ and normalization such that the spectra match at E$ > 10$~keV. 

Simply by eye it is clear that these spectra strongly deviate from a power-law by softening above 10\,keV. We emphasize that the effect of PSF mixing is to obscure the true underlying spectra. It cannot create the appearance of a broken power-law out of a superposition of spectra if they were solely power-laws. 

We present now the spatially resolved maps. Figure \ref{contour1} and Figure \ref{contour2} show maps of the (1) the break energy, (2) $\Delta\Gamma = \Gamma_2 - \Gamma_1$), (3) $\Gamma_{<6keV}$ (the photon index at energies $<6$keV), and (4) $\Gamma_2$. In general the errors in individual boxes are ($\Gamma_1, \Gamma_2) \sim \pm 0.015$ and the error in the break energy (if there is one) $\pm$1\,keV. The left panels show the map overlaid with \textit{NuSTAR} intensity contours and right panel with \textit{Chandra} contours to show details of the spatial structures. The pulsar is represented by a cross and located in all images at $(\Delta$Ra,$\Delta$Dec) = (0,0). Although we show the map for FPMA for ease of presentation, the maps from FPMB are identical within errors.

Figure \ref{contour1} shows that inside the remnant the break energy is roughly constant, with an average value of $\sim$9\,keV decreasing towards the NW and increasing at the SE edge. The $\Delta\Gamma$ map shows that the highest values of $\Delta\Gamma$ follows the curvature of the forward edge of the torus. It is interesting that the largest value is not found at the location of the highest intensity, but rather slightly more north on the edge of the \textit{Chandra} intensity contour. Comparing to the break energy map, where $\Delta\Gamma$ is high the break energy is on average lower.

Figure \ref{contour2} shows the low-energy spectral index, $\Gamma_{<6keV}$. The 6~keV energy was chosen because it is always below the break energy. The greatest change in morphology in this map occurs along the forward edge of the torus where the spectrum softens rapidly. Along a line from the pulsar towards the NW corner, the radial spectrum softens faster above the break than below. This holds at the majority of azimuthal angles around the torus, but not in the jet direction. Neither the jet region or counter-jet region have measurable spectral breaks. The SE corner appears to have a negative $\Delta\Gamma$, but this is likely the effect of the PSF scattering the harder torus high energy spectrum into the softer nebula.

There is no easy way to spatially disentangle spectral components that have been mixed by the PSF (In the Appendix we provide details on the \textit{NuSTAR} PSF and discuss the technical details of it). To interpret the spectral map, we therefore compose a model of the nebula and forward fold it through the \textit{NuSTAR} response and compare it to the data. We create a 2-D spectral model based on the analysis of \citet{Mori2004} of the Crab with \textit{Chandra}. This data set shows spectral variations on arcsecond scales, but we follow the authors comment that the remnant can be represented by six components: (1) a halo (shown in dark red in Figure \ref{contour3}) with $\Gamma = 3.0$, (2) a cap (yellow) $\Gamma=2.5$, (3) a skirt (light blue) $\Gamma = 2.1$, (4) a torus (dark blue) $\Gamma = 1.9$, (5) a jet (medium blue) $\Gamma = 2.0$, and (6) a center (black) $\Gamma = 1.6$. We assume that this spectral model holds up until our measured break. Based on the \textit{NuSTAR} analysis, we further assume that the torus has an average spectral break at 9\,keV with a $\Delta\Gamma=0.25$, and that the spectrum of the center has a break at the same energy, but with a $\Delta\Gamma=0.1$. We arrived at this smaller $\Delta\Gamma=0.1$ through a series of iterations since this component is maximally obscured and difficult to determine from the \textit{NuSTAR} data. The center and the torus are the only two components that we allow to have a break. All other regions are presumed to have power-law spectra. We used a \textit{Chandra} image of the Crab, shown in Figure \ref{contour3}, to set the normalization of the map in the  2 -- 10\,keV band. The spectra are completely defined by this set of parameters.

We propagated this model through the \textit{NuSTAR} PSF and responses and analyzed the output in exactly the same manner as the real data itself. Figure \ref{contour4} shows the resulting maps. Macroscopically they reproduce the {\em NuSTAR} observations very well. There are some discrepancies, but these can be explained by the simple model we use which does not include the low energy variations observed by \citet{Mori2004} or the break energy variations seen by \textit{NuSTAR}. However, these simulations still address some important questions. It is clear that the torus spectrum has to steepen with energy, and that one has to use an approximately constant spectral index in both $\Gamma_1$ and $\Gamma_2$ all the way out to the edge, where there is a rapid transition to a softer, unbroken spectra. We tested this hypothesis by creating a model with a linearly changing spectral index as a function of radius as has been observed for G21.5-0.9 \citep{Safi2001} and 3C\,58 \citep{Slane2004}. The resulting maps do not match the observations and create spectral slopes that are far too soft in the interior. The spectrum of the core must be hard in order to reproduce the \textit{NuSTAR} results. \citet{Mori2004} did not find a hard center of the remnant below 10 keV, but their observations suffered pile-up in the center, and they couldn't measure the spectrum in the inner ~10\as. The origin of the hard component seen by {\em NuSTAR} could be the neutron star itself, which according to \citet{Weisskopf2011} has a spectral index of $\Gamma\sim1.9\pm0.4$. Because the simulations are driven by the \textit{Chandra} flux map, the location of the pulsar has no counts because it was excised, which may explain why instead of a point source our simulations required an extended, hard core, distributing the missing flux over a large region. Finally neither the cap or the jet appear to steepen with energy

It is pertinent to question whether dust scattering could have anything to do with the slope below 10\,keV.  Unless the N$_H$ column inside the remnant is larger than the galactic column by a factor of 10, the interstellar dust extinction has little effect above 3\,keV \citep{Seward2006}. Indeed the fact that the break energy and $\Delta\Gamma$ trace out features of the remnant strongly indicates that it must be intrinsic to the source.

\subsection{Spatial Extent of the Nebula}\label{size}

\begin{figure*}
\begin{center}
\includegraphics[width=1.0\textwidth]{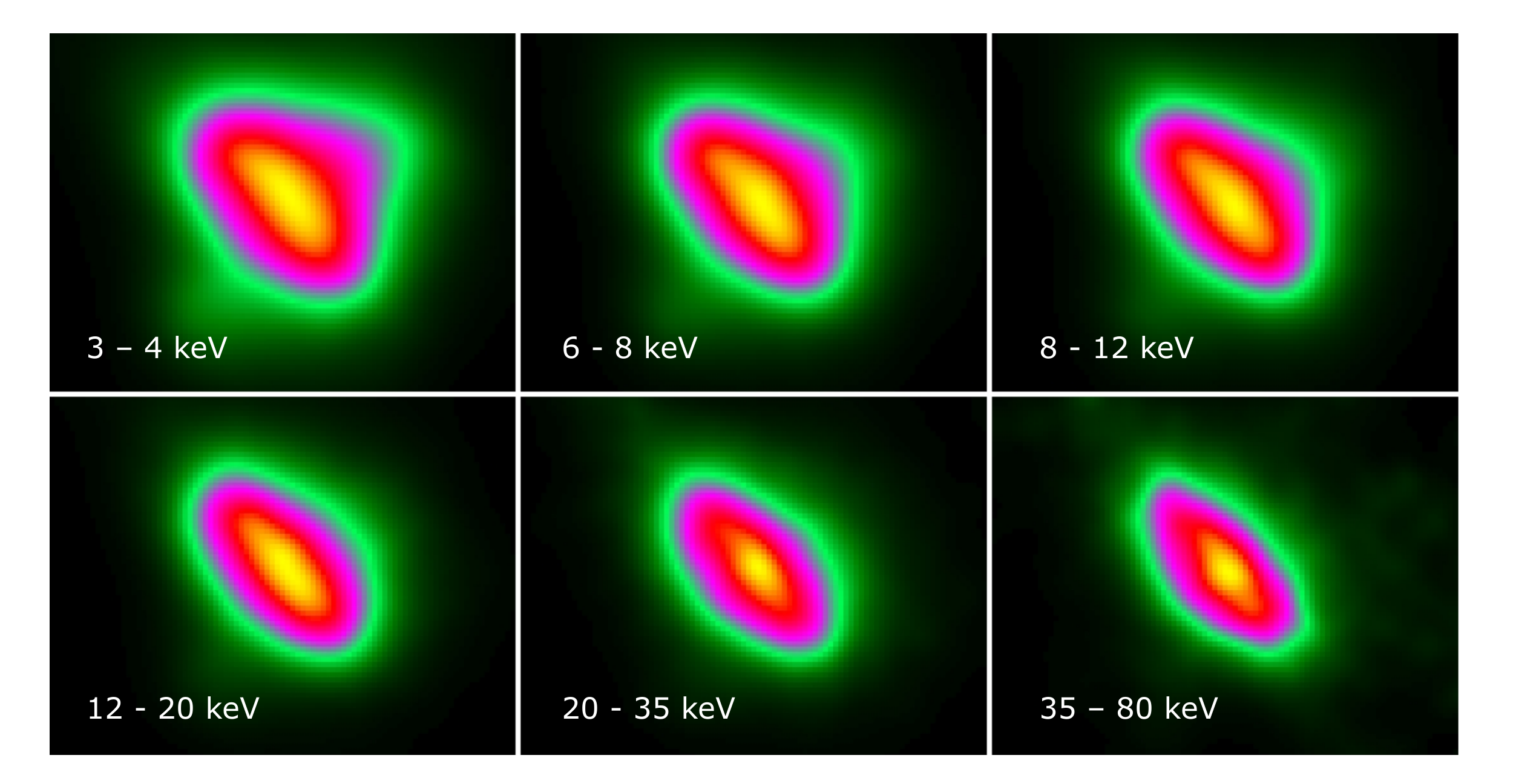}
\end{center}
\caption{Maximum likelihood deconvolved images of FPMA in six different energy bands shown in a square root stretch during pulse off. The shrinking of the NW counter-jet is easily observable and the morphological changes of the NE torus can also be seen.}
\label{deconvolved}
\end{figure*}

The size of the Crab remnant shrinks as a function of energy due to the radiative lifetimes of outward-propagating electrons being shorter for high energy than low energy particles. This effect is often referred to as `synchrotron burn-off'.  To investigate the radial extent of the Crab as a function of energy we deconvolved the {\em NuSTAR} maps using a maximum likelihood method. The deconvolution procedure is sensitive to artifacts, such as detector gaps and the variation in signal to noise ratio with position in the map. The stronger the source relative to background the better the deconvolution results. The  PSF is relatively constant near the optical axis (the difference in the HPD between off-axis angles of 1\am\ and 2\am\ is less than 1\as), but becomes azimuthally distorted at large off-axis angles. This, however, does not become noticeable until about 3\am\ off-axis, where the difference between the major and minor axis of the PSF is $\sim$2\%. To minimize these effects we selected a subset of the observations in Table \ref{obsid} marked with ``b'' at off-axis angles less than 2\am\ and well away from the detector gaps. At these off-axis angles we can design an average PSF weighted by time and combine the images to yield a more robust result than individually deconvolving short segments. 

To remove the contamination of the pulsar we only use photons falling in phase bins 10 -- 12 and deconvolve the Crab in the following energy bands: 3 -- 5, 5 -- 6, 6 -- 8, 8 -- 12, 12 -- 20, 20 -- 35, and 35 -- 78\,keV. Prior to combining the images, we vignetting correct them with the effective area taken at the area-weighted average energy.

Deconvolution with a maximum likelihood method is iterative and if not performed with care can introduce artifacts due to over deconvolution. During the deconvolution process we checked the relative size between the selected iteration steps  (20, 30, 40 and 50) within each energy band, and saw that the size difference remained largely unchanged after 30 deconvolutions. For the highest energy band, however, we started to observe artifacts after 50 deconvolutions, and we therefore estimate that 40 deconvolutions is a safe number to use for all energy bands.

Because of the nature of the deconvolution process, it is difficult to assign an error to the resulting size. Ideally a model of the source can be forward folded and compared to the actual image, but in the case of the Crab where such a model is not well known, and the morphology of the source is complex, this approach is not feasible. Forward folding the deconvolved image with the PSF and comparing to the raw image is another option, but it only serves to identify gross errors. It is therefore not possible to assign an error to the absolute size in the deconvolved images. Fortunately, we are interested in a rate of change as a function of energy, and we can investigated the error of the deconvolution as a function of energy by deconvolving several strong \textit{NuSTAR} point sources. These should have a constant radial extent as a function of energy, and any discrepancy is assumed to be the error introduced by deconvolution. We used the same energy bands as for the Crab images, and found an evolution in the deconvoled point source sizes as a function of energy. The discrepancy between the highest and lowest energy bands is $\sim$1.5\as, and we have conservatively assumed a 2\as\ 1$\sigma$ error in the relative size of the remnant between the energy bands.

\begin{figure}
\begin{center}
\includegraphics[width=0.5\textwidth,viewport=0 0 570 570]{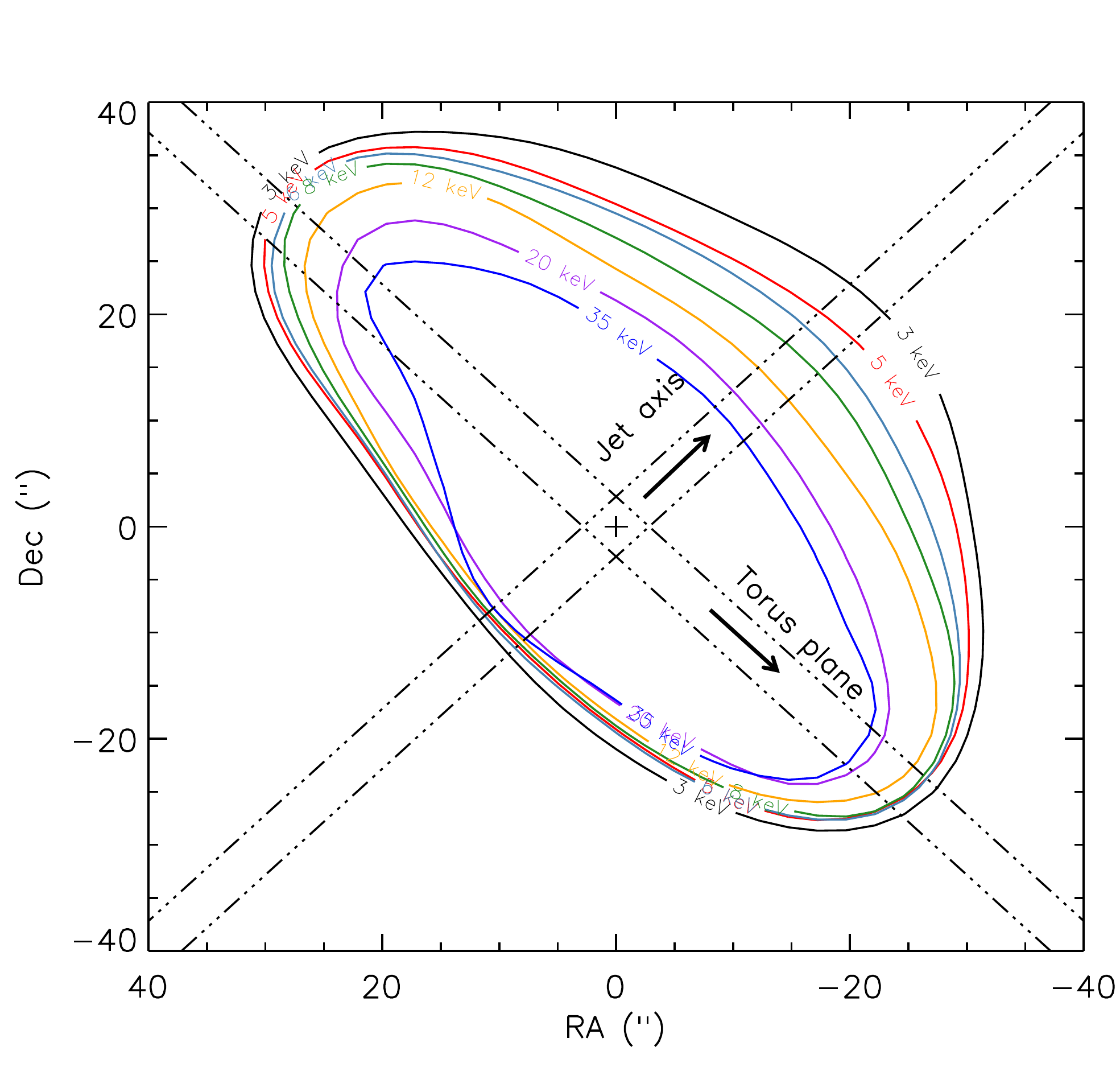}
\end{center}
\caption{Image of the HWHM contours for energy bands: 3 -- 5, 5 -- 6, 6 -- 8, 8 -- 12, 12 -- 20, 20 -- 35, and 35 -- 78\,keV.}
\label{sizecontour}
\end{figure}

\begin{figure}
\begin{center}
\includegraphics[width=0.5\textwidth,viewport=50 200 550 620]{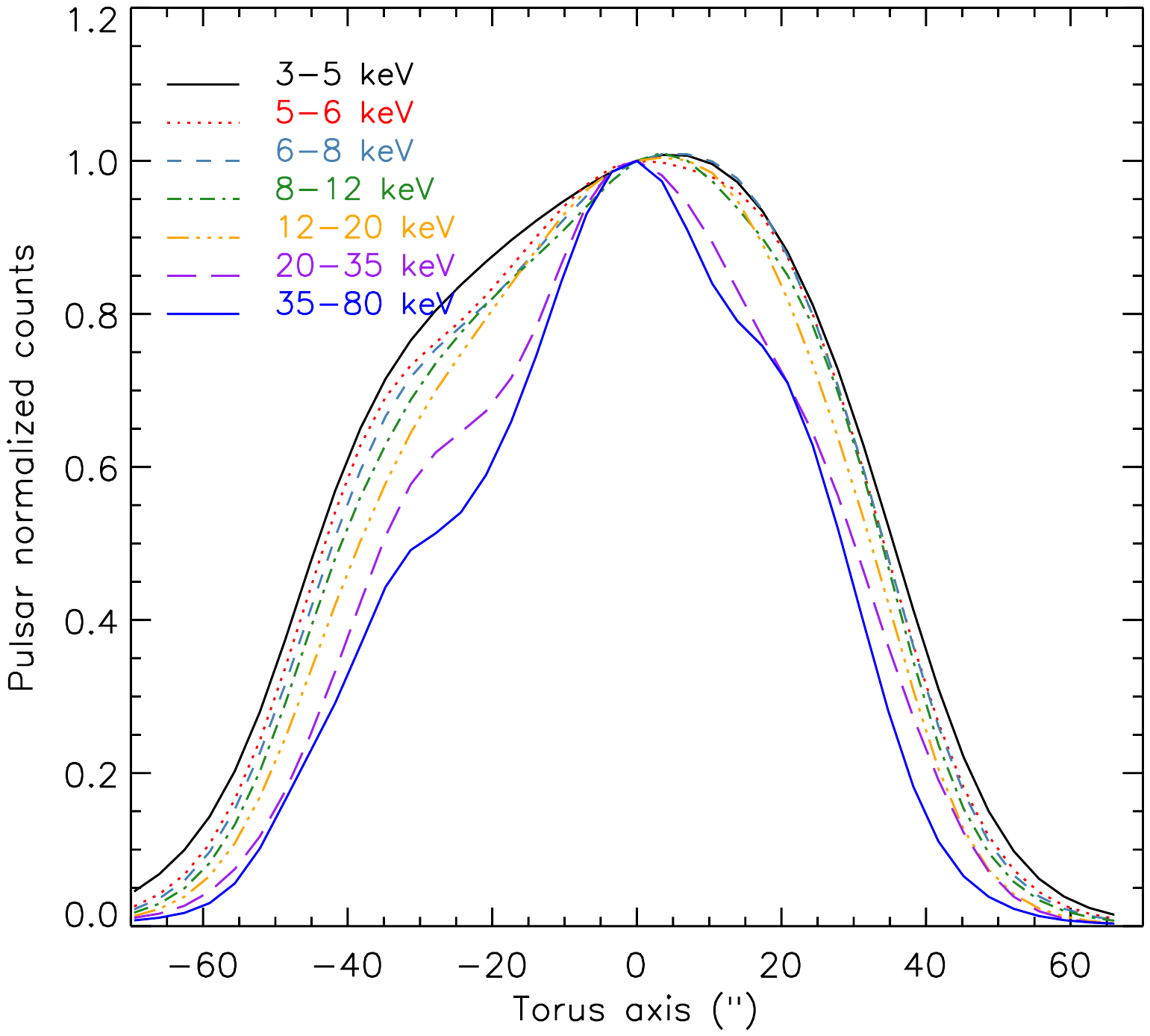}
\includegraphics[width=0.5\textwidth,viewport=50 200 550 620]{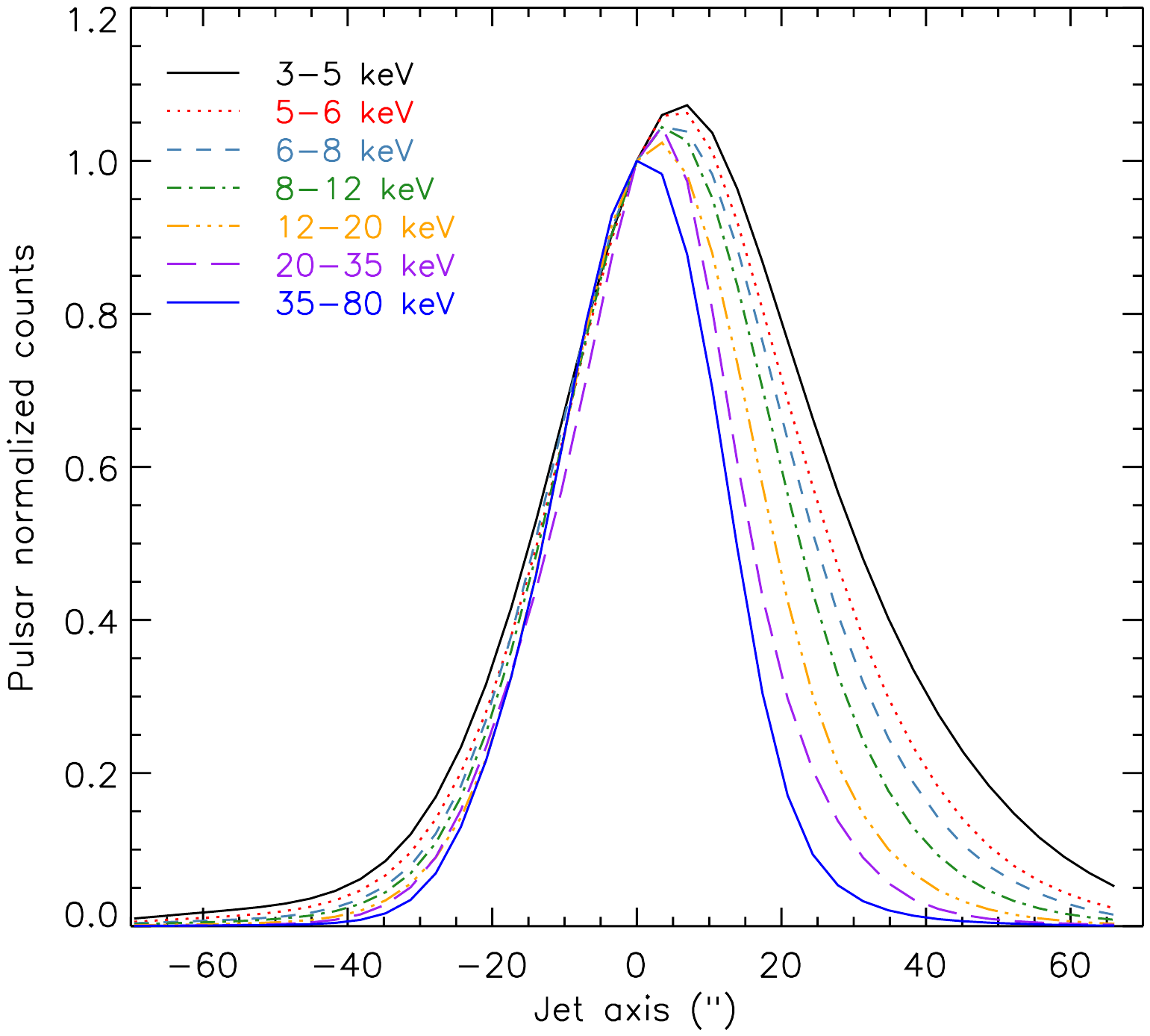}
\end{center}
\caption{Energy dependent profiles of module A interpolated along the torus plane and jet axis of the deconvolved images shown in Figure \ref{sizecontour} in the direction of the arrows. Top: Profile along torus plane. The right shoulder corresponds to the SW. Bottom: Profile along jet axis. The right shoulder corresponds to the NW counter-jet and shows a very clear shrinking as a function of energy, while the left shoulder corresponds to the SE jet and shows almost none at all.}
\label{deconvprofile}
\end{figure}

Figure \ref{deconvolved} shows the resulting images for FPMA. The major components observed in \textit{Chandra} are also seen by \textit{NuSTAR}, and morphological changes as a function of energy are clearly visible. Figure \ref{sizecontour} plots the contours of the HWHM as defined by the point where the intensity falls to half the value measured at the pulsar location, and shows the magnitude of the shrinkage is position angle (PA) dependent. Figure \ref{deconvprofile} shows the profiles of two perpendicular slices with the profiles extracted by bi-linear interpolation. The intensity map is off-set from the pulsar due to the 27$^\circ$ torus inclination and the beaming of the leading edge of the torus, but we normalized the curves at the pulsar location, which places the peak intensity towards the NW for low energies. The offset decreases with increasing energy and is gone by 20 keV. The NE and SW side of the torus plane therefor have somewhat different morphologies close to the pulsar, and it can also be seen that the jet and counter-jet directions evolve quite differently; the counter-jet and cap area HWHM falls off rapidly with energy, while the forward jet is much slower.

\begin{figure}
\begin{center}
\includegraphics[width=0.55\textwidth,viewport=100 350 600 650]{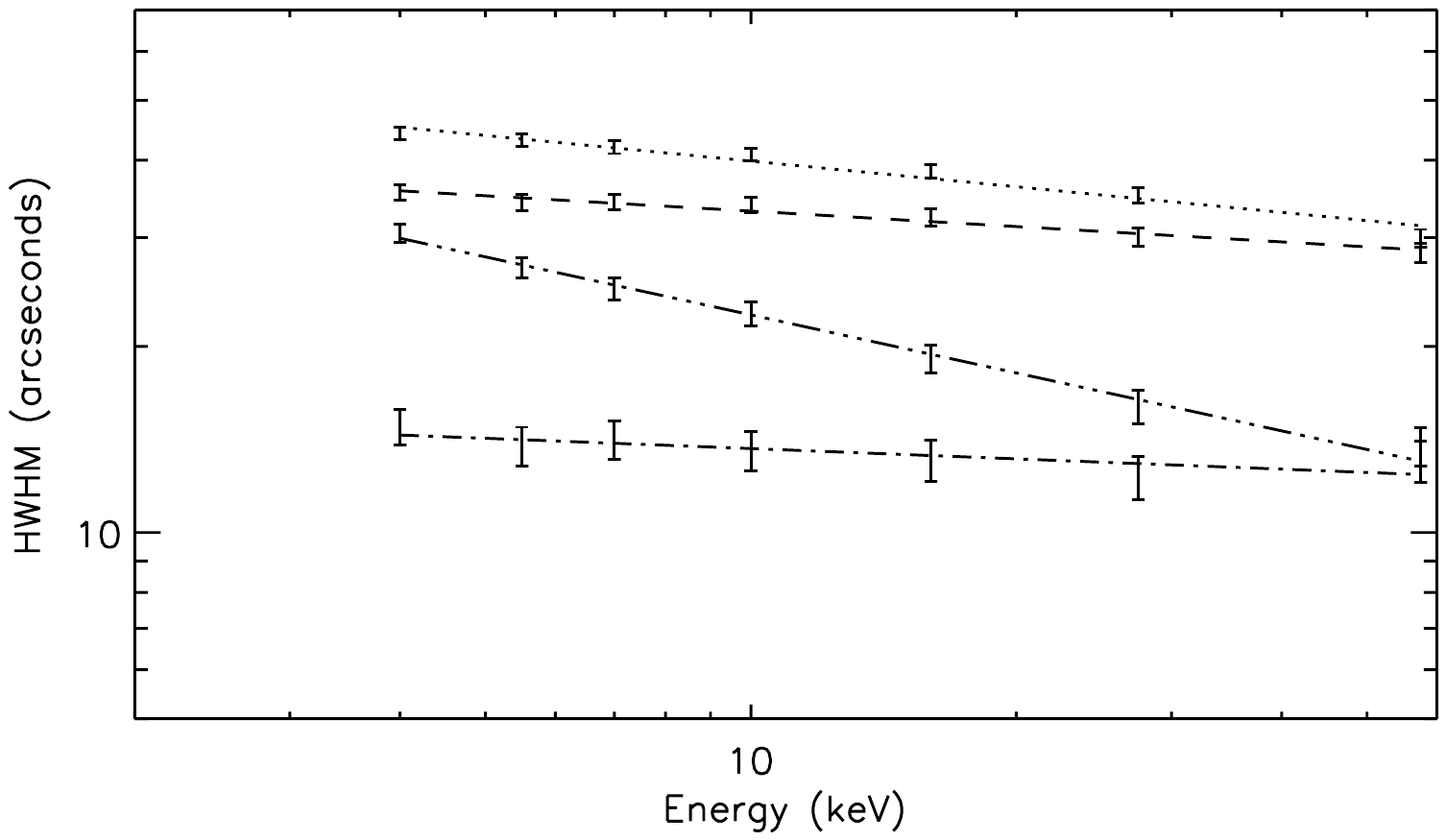}
\end{center}
\caption{Half Width Half Max, HWHM, in arcseconds. The HWHM is measured from the pulsar location along the torus and jet axis for either side. From top to bottom: NE torus plane. SW torus plane. NW jet axis. SE jet axis.}
\label{fits}
\end{figure}

We fitted the HWHM of the profiles as a function of energy using a power-law $kE^{-\gamma}$. Figure \ref{fits} shows the fits for FPMA for the two sides of both the jet axis and torus plane, and Table \ref{tablefits} summarizes the fit values for both FPMA and FPMB. The flux of the jet is much smaller than that of the torus, and what we measure along the SE is most likely not the jet, but rather the edge of the torus. The rate of change in NE, SE and SW are of a similar slope, but different intensity, indicating the shrinkage occurs in the plane of the torus. The projection effect of the tilted plane would in this way explain the smaller magnitude observed along the SE edge. The NW edge of the torus has a line of sight that also includes the NW cap and counter-jet, obscuring the true shrinkage of the torus in this direction. We know this area has the softest spectrum and the index rate of shrinkage, $\gamma$, is twice as large as that of the torus.


\begin{table}
\centering
\caption{Nebula HWHM}
\begin{tabular}{ccc|cc}
\hline
&\multicolumn{2}{c}{FPMA}&\multicolumn{2}{c}{FPMB} \\
\hline
Axis & $\gamma$ & k & $\gamma$ & k\\
\hline
NE Torus & 0.094$\pm$0.018 & 49(1) & 0.079$\pm$0.017 & 50(1)\\
SW Torus & 0.060$\pm$0.020 & 38(1) & 0.085$\pm$0.02 & 42(1)\\
SE Jet & 0.083$\pm$0.062 & 13(1) & 0.014$\pm$0.056 & 12(1)\\
NW Jet & 0.245$\pm$0.029 & 46(2) & 0.192 $\pm$0.027 & 42(2)\\
\hline
\label{tablefits}
\end{tabular}
\end{table}

\section{Discussion and Conclusion}

\citet{Gratton1972} and \citet{Wilson19722} developed the first analytic models attempting to explain the observational properties of the Crab. They constructed a model of the nebula assuming that electrons and positrons produced by the pulsar are transported away by diffusion, with energy lost through synchrotron radiation. The size and spectral shape of the nebula in the radio and optical are well explained by the model, but the authors did not consider the details of the production of the electrons and positrons by the pulsar, or their relationship to the pulsar properties, such as its magnetic field. 
 
\citet{Rees1974} and \citet{KC19841, KC19842} constructed a steady state spherically symmetric magnetohydrodynamic (MHD) model of the Crab with a toroidal magnetic field, which links the pulsar to the nebula. A highly relativistic pulsar wind is terminated by a strong MHD shock, which decelerates the flow to non-relativistic speeds.  Downstream the approximately adiabatic flow continues to decelerate, carrying its frozen-in magnetic field out to the supernova ejecta that confines the synchrotron nebula. The model successfully describes the integrated spectrum from optical to X-rays where the diffusion model fails. However, it does not reproduce the radio spectrum, nor reproduce the spatial variations of the optical spectrum across the remnant. The model predicts a constant spectrum that steepens only at the edge of the remnant, but this has not been met by observations in the optical, which instead show a monotonically increasing spectral index \citep{Veron1993, Temim2006}. Similar discrepancies have been observed in X-rays for the other two young PWN, 3C\,58 and G21.5-0.9 \citep{Nynka2014,Slane2004,Safi2001}. This motivated \citet{Tang2012} to re-investigate the diffusion driven models from \citet{Gratton1972} and \citet{Wilson19722}. They found that a combination of diffusion and advection can reproduce the observed spectral index variations observed in radio and optical for the Crab, and in X-rays for the two other sources. However, diffusion under-predicts the size of the Crab in X-rays. A combination of advection and diffusion increases the predicted half-light radius of the Crab at 10 keV from 20\as\ in the pure diffusion case to 25\as\ for the best fit combination of advection and diffusion. This is still too small, and the conclusion of \citet{Tang2012} is that due to the compact size of the Crab in X-rays, the advection time scale dominates over the diffusion time scale for X-ray producing particles.  Detailed \textit{Chandra} maps of the Crab nebula by \citet{Mori2004} have shown that the spectral index, within some small-scale variations, is constant for most of the nebula and steepens abruptly at the edge. This supports the idea of an advection, rather than diffusion, dominated X-ray nebula. 

The first measurements of the extent of the X-ray Crab as a function of energy from 2 -- 12~keV were made in 1964 using a lunar occultation method \citep{Bowyer1964}. \citet{Ku1976} later combined the results of several experiments \citep{Palmieri1975, Ricker1975, Fukada1976, Ku1976}, and found the size to vary as $\propto\nu^{-\gamma}$ with $\gamma = 0.148 \pm 0.012$, an effect they attributed to a combination of diffusive and advective transport of electrons in the nebula.  At this time the advective model of KC84 had not yet been constructed, but KC84 concluded in their paper that these results agree fairly well with an advective electron transport. These rates were obtained from multiple experiments and observatories along a narrow range of position angles through the Crab, and they did not probe any azimuthally dependent information.

With \textit{NuSTAR} we have measured the energy dependent size at all position angles at energies from 3 -- 78\,keV. The spatial dependence of the HWHM along the edge of the torus plane is well fit by a power-law with an average of $\gamma=0.08 \pm 0.03$. This is in good agreement with the predictions made by \citet{KC19842} of $\gamma \sim 1/9$. Measurement of the jet rate of change could not be done since the torus flux dominates, but judging from the tail of the profile in Figure \ref{deconvprofile} bottom panel, it does appear as if the jet shrinks more swiftly outside the torus as a function of energy. The size of the counter-jet as a function of energy clearly follows a different rate, $\gamma=0.22\pm0.03$, suggesting the energy transport is not the same as in the torus. In other band passes this region is found to be different as well, and is the only part of the remnant where the synchotron nebula extends beyond the edge of the visible filaments. It is purported that here the shock  velocity is larger than for the rest of the remnant, and the post-shock cooling time longer than the age of the Crab \citep{Sankrit1997}.

Similar values of $\gamma \sim 0.2$, have been found for the nebula remnants in MSH\,15--52 \citep{An2014} and G21.5-0.9 \citep{Nynka2014}, two PWN observed with \textit{NuSTAR} and covering the same energy band. The authors conclude that both remnants appear to be diffusion dominated. Both of these remnants are more than twice as distant ($\sim$5 kpc) and have distinct morphological differences from the Crab. MSH\,15--52 is not a spherically symmetric remnant, but highly irregular and dominated by a powerful jet. G21.5-0.9 is spherically symmetric, but the inner portions of the remnant are not resolved and there is no apparent torus or jet evident. This would suggest that for the Crab we are probing the inner workings of the PWN, namely the torus where advection processes dominate, while in the counter-jet region diffusion processes dominate similar to what is found in MSH\,15--52 and G21.5-0.9.

Using a map of the average spectral index distribution from \textit{Chandra} as the basis, we have shown through simulations that the torus must have a near constant spectrum as a function of radius. If we presume that the wind expands radially along the torus plane, then the transition from a flat profile to a steep occurs at a radius of approximately at 50\as. This is consistent with the KC84 model, which predicts a flat profile in an advection dominated MHD flow that transitions at $\sim 50$\as\ if the termination shock is of size 10\as. In this geometry, the energy dependent rate of shrinkage we have found is also consistent with KC84. We confirm that the torus is well-described by a portion of a spherical outflow, with relatively little diffusion or turbulent mixing and conclude that for the X-ray band at least, advection dominates the processes. 

What was not anticipated is that in the \textit{NuSTAR} band there is a very clear spectral steepening inside the remnant. Our analysis has revealed the spatially resolved nebula is not well fit by a power-law across the band 3 -- 78~keV, but rather needs a model with a spectral steepening. We chose a broken power-law as an adequate representation and presented maps of the spectral variation that show a steepening of $\Delta\Gamma \sim 0.25$, which appears to be limited to the torus of the nebula with an approximate break energy at $\sim9$\,keV. We confirm this with simulations in order to address any ambiguities resulting from spatial smearing by the PSF. We also conclude that the jet and counter-jet regions do not appear to have a similar spectral steepening.

The integrated spectrum of the nebula has been known to turn over above 100\,keV, transitioning from a $\Gamma \sim 2.1$ to $\sim$ 2.14. The gradual break must come about from the shrinking of the nebula, and the steepening of the torus spectrum. There is no immediate physical interpretation of the torus spectrum. It is possible the steepening could result from the projections of the emission from electron populations of different synchrotron ages, but a similar steepening has also been by \textit{NuSTAR} in the spatially integrated spectrum of G21.5-0.9 \citep{Nynka2014}, and since the two remnants are morphologically different, it questions the geometrical argument and might suggest the steepening is connected to the injection spectrum itself. In either case the steepening of the torus spectrum shouldn't come as a surprise. Hard X-ray instruments have measured the photon index above 100\,keV to be $2.140 \pm 0.001$ \citep{Pravdo1997} while \textit{Chandra} measures the average photon index of the torus to be $\sim$1.9 \citep{Mori2004}. This is a softening of $\Delta\Gamma \sim 0.25$, which matches nicely with the number found in our maps. Using the rate of burn-off for the NE side from Table \ref{tablefits}, the HWHM radius of the Crab is just $\sim$30\as\ at 100\,keV, restricting the source of the steepening to come from the innermost regions of the remnant. It is therefore not only likely but necessary for the torus to steepen in the \textit{NuSTAR} band in order to bridge the gap between soft X-ray and $\gamma$-ray observation.

We performed phase resolved spectroscopy of the Crab on several different length scales from the inner 17\as\ out to 200\as. The pulsed spectrum is best represented by a steepened spectrum with a break energy of $\sim$10\,keV. As found previously by \citet{Weisskopf2011} and \citet{Willingale2001}, the index below 10\,keV, $\Gamma_1$, shows spectral evolution as a function of phase; the secondary pulse is harder than the primary and the hardest index occurs during the bridge emission between the first and secondary pulse. We find that $\Gamma_2$ traces $\Gamma_1$ with an average $\Delta\Gamma = 0.27 \pm 0.09$. There are indications of this steepening in RXTE \citep{Pravdo1997} observations, but it was not properly quantified. \citet{Kuiper2001} present phase resolved pulsed spectra by combining \textit{BeppoSAX}, \textit{CGRO}, and \textit{GRIS} data and found that from 0.1\,keV up to 10\,GeV it could be fit using three components: (1) a power-law, (2) a modified power-law (\texttt{logpar}) for the first pulse, and (3) a modified power-law for the bridge emission. They found acceptable fits by combining these three models, and we attempted to fit with the same combination. While we were able to find statistically acceptable fits, the models proved degenerate in our narrower band-pass without the $\gamma$-ray spectrum to constrain them.


Finally \textit{NuSTAR} participated in a ToO observation of the Crab during the flaring on the 9th of March 2013. No flux change was detected between 3 -- 78 keV aside from what is expected from calibration, which is $\pm 5$\% in flux and $0.01$ in spectral index change, and thus places an upper limit on the hard X-ray variability due to $\gamma$-ray flares. 

\acknowledgments
This work was supported under NASA Contract No.
NNG08FD60C, and made use of data from the NuSTAR mission,
a project led by the California Institute of Technology,
managed by the Jet Propulsion Laboratory, and funded by the
National Aeronautics and Space Administration. We thank
the NuSTAR Operations, Software and Calibration teams for
support with the execution and analysis of these observations.
This research has made use of the NuSTAR Data Analysis
Software (NuSTARDAS) jointly developed by the ASI Science
Data Center (ASDC, Italy) and the California Institute
of Technology (USA).

\section{Appendix}
The \textit{NuSTAR} point spread function has a sharp core (FWHM = 18\as) but extended large wings. It has an energy dependency, which causes the half power diameter to shrink as a function of energy by a few arcseconds between 3 and 10 keV. Above 10 keV the PSF remains constant. 

\begin{figure}[t]
\begin{center}
\includegraphics[width=0.47\textwidth]{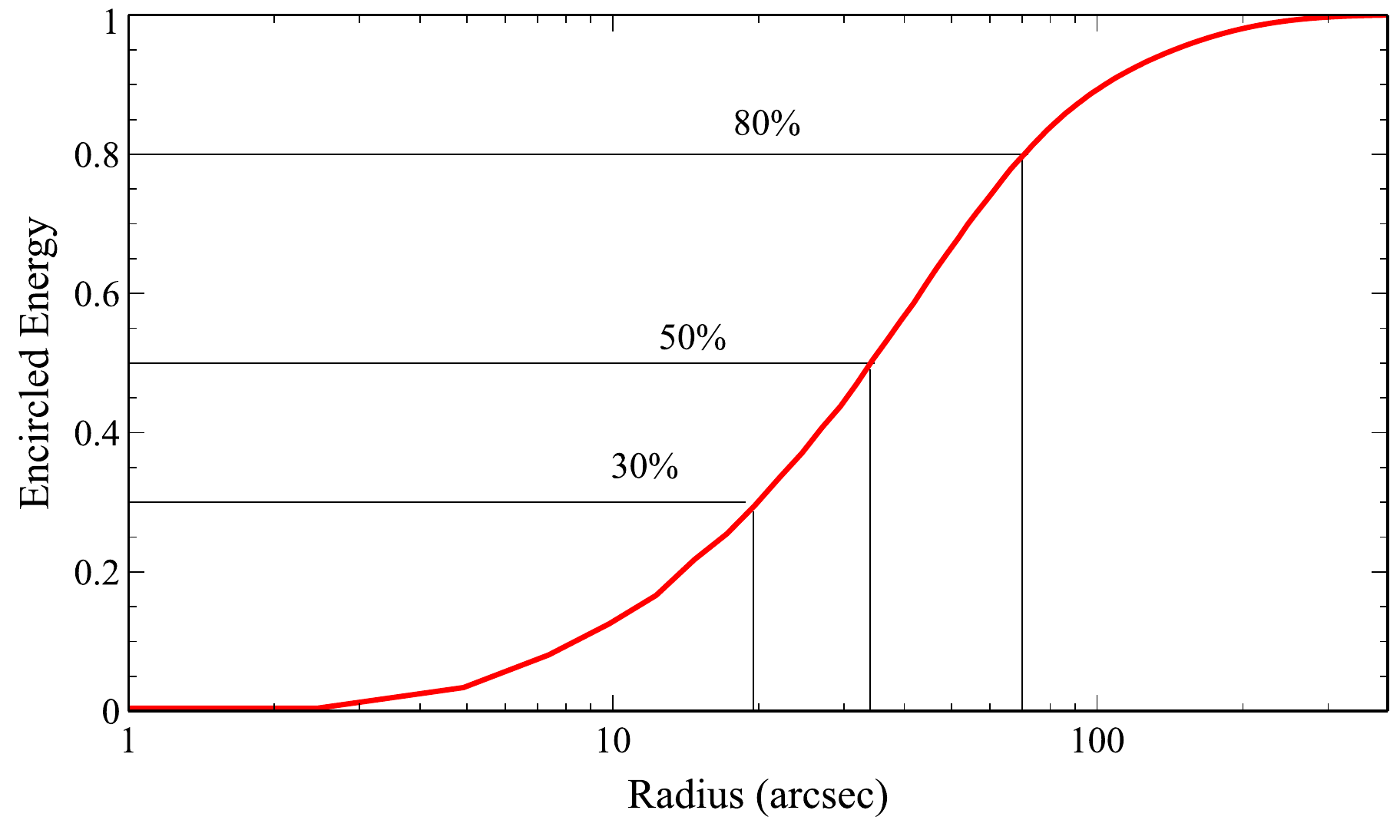}
\end{center}
\caption{\textit{NuSTAR} encircled energy curve at 10 keV.}
\label{eefplot}
\end{figure}

As shown in Figure \ref{eefplot} of an encircled power curve of a 10 keV PSF, 30\% of the photons of a point source are located within a radius of 20\as, 50\% within 35\as\ and 80\% within 70\as. This means that for an extended source like the Crab, which has most of its flux contained within an ellipse of 100\as$\times$150\as, the extended wings will roughly redistribute 60\% of the flux from a central volume element over the rest of remnant. This causes the true spectral distribution to be mixed, and even with detailed knowledge of the PSF and effective area, it is unfortunately not possible to disentangle these mixed spectral distributions by any method of deconvolution. 

One approach to tackle this mixing, is to subdivide the extended source into smaller regions and calculate the cross-correlation functions between the regions of both the ARFs and PSFs as was done in \citet{Wik2014}. The resulting spectra and responses must then be fitted simultaneously. This works well for large sources with spectra that are slowly changing, but for for smaller sources with fast changing spectra where the number of regions could number in the hundreds, this starts getting cumbersome. Another way is to forward fold a two dimensional spectral model through the optics response, carefully taking into account the PSF, and compare to actual data.

Based on the fine structure we observe see in the spectral data maps, we have in chosen the latter method.

\bibliography{bib}

\end{document}